\documentclass[10pt,journal,compsoc,onecolumn]{IEEEtran}
\usepackage{url}
\usepackage{amsfonts}
\usepackage{amsmath}
\usepackage{amssymb}
\usepackage{color}
\usepackage{graphicx} 
\usepackage{epstopdf}
\usepackage{bm}
  \usepackage{amsthm}
\usepackage{dcolumn}
\usepackage{bm}
\usepackage{hyperref}
\usepackage{dsfont}
\usepackage[caption=false]{subfig}
\usepackage{float}
\usepackage{microtype}
\usepackage{lipsum}
\usepackage{cuted}

\begin{document}
\title{Assessing Dengue Risk Globally Using Non-Markovian Models}

\author{Aram Vajdi$^{*}$ , Lee W. Cohnstaedt$^{1,\ddagger}$ and Caterina M. Scoglio$^{*}$

\IEEEcompsocitemizethanks{\IEEEcompsocthanksitem $^{1}$ Corresponding author 
\protect\\
$^{*}$ Department of Electrical and Computer Engineering, Kansas State University,
Manhattan, KS 
\protect\\ $^{\ddagger}$ United States Department of Agriculture, Agricultural Research Service, Foreign Arthropod-Borne Animal Diseases Research Unit, Manhattan, Kansas
}     
}

\IEEEtitleabstractindextext{%
\begin{abstract}
Dengue is a vector-borne disease transmitted by \textit{Aedes} mosquitoes. The worldwide spread of these mosquitoes and the increasing disease burden have emphasized the need for a spatio-temporal risk map capable of assessing dengue outbreak conditions and quantifying the outbreak risk. Given that the life cycle of \textit{Aedes} mosquitoes is strongly influenced by habitat temperature, numerous studies have utilized temperature-dependent development rates of these mosquitoes to construct virus transmission and outbreak risk models. In this study, we advance existing research by developing a mechanistic model for the mosquito life cycle that accurately accounts for the non-Markovian nature of the process. By fitting the model to data on human dengue cases, we estimate several model parameters, allowing the development of a global spatiotemporal dengue risk map. This risk model employs temperature and precipitation data to assess the environmental suitability for dengue outbreaks in a given area. Furthermore, we demonstrate how to reduce the model to the corresponding differential equations, enabling us to utilize existing methods for analyzing the system and fitting the model to observations. This approach can be further applied to similar non-Markovian processes that are currently described with less accurate Markovian models.     
\end{abstract}

\begin{IEEEkeywords}
 Non-Markovian Models, Dengue, \textit{Aedes} mosquitoes
\end{IEEEkeywords}}

\maketitle

\IEEEdisplaynontitleabstractindextext

%
\IEEEpeerreviewmaketitle

\IEEEraisesectionheading{\section{Introduction}\label{sec:introduction}}
Dengue fever, a vector-borne viral disease transmitted primarily by \textit{Aedes} mosquitoes, presents a significant global public health challenge \cite{bhatt2013global}. The main carrier of the dengue virus is the \textit{Aedes aegypti} mosquito, an urban species whose life cycle is strongly influenced by environmental temperature. Additionally, \textit{Aedes aegypti} serves as the primary vector for the Chikungunya, Zika, and yellow fever viruses \cite{mordecai2017detecting,scott2012feeding,messina2016mapping}. The global prevalence of \textit{Aedes} mosquitoes, along with its expansion into new regions \cite{heydari2018dengue}, necessitates the development of outbreak risk models which are crucial for guiding the control strategies aimed at eliminating virus transmission.

In recent years, there has been a growing interest in developing mechanistic models that integrate entomological and environmental components to gain a better understanding of the spread of dengue \cite{pinho2010modelling,yang2011follow,yang2009assessing,yang2009assessing2,brady2014global,
mordecai2017detecting,chen2012modeling,yang2017transovarial,
pliego2017seasonality,rossi2014modelling,otero2008stochastic}. For instance, Yang et al. conducted a study in which they measured entomological parameters of \textit{Aedes aegypti}, including temperature-dependent development and mortality rates \cite{yang2011follow}. They incorporated these parameters into differential equations that describe dengue transmission within a transmission model. Using this mechanistic model, they determined that a temperature of $29^{\circ}\mathrm{C}$ is the most favorable for a dengue outbreak \cite{yang2009assessing,yang2009assessing2}. In another study \cite{mordecai2017detecting}, the authors utilized a calibrated basic reproductive number for dengue to develop an outbreak risk model. In the field of mathematical epidemiology, the reproductive number is defined as the number of secondary infected cases generated by one infected case in a healthy population. Therefore, if this number is greater than one, the number of new cases increases. Nevertheless, it's crucial to highlight that even though both of these studies incorporate temperature-dependent biological parameters, their mathematical equations and subsequent derivations rest upon the assumption of a constant environmental temperature. In another noteworthy study \cite{kraemer2015global}, researchers compiled a global database of \textit{Aedes} mosquitoes occurrences and employed pertinent environmental variables—such as temperature susceptibility, precipitation, and urbanization—as predictive factors to develop a niche model that defines the global suitability map for \textit{Aedes} mosquitoes presence.

In the majority of existing mechanistic models for mosquito-borne diseases, the mosquito life cycle is divided into multiple phases, including aquatic stages and the adult stage. Subsequently, by leveraging experimental data on developmental and mortality rates at each of these stages, a set of differential equations is formulated to provide a mathematical description of the mosquito population dynamics. For example, using parameters $\alpha$ and $\beta$ to represent the development rates of mosquito larvae and pupae, and $\mu$ to denote the mortality rate of pupae, the following equation: 
\[
\frac{dp}{dt}=\alpha l-(\beta+\mu)p
\]
has been employed to estimate the mosquito pupae population represented by variable $p$, with the larvae population being denoted as $l$. However, even in an environment with a constant temperature and consequently constant development and mortality rates, such an equation does not precisely describe the mosquito life cycle. This discrepancy arises from the fact that this equation describes Markovian transitions with exponential sojourn times, while in reality, this is not always the case. In fact, a new pupa emerges when the larval development is complete, and this development is not necessarily exponentially distributed; it often occurs around a specific time that depends on the temperature. Therefore, the actual process is non-Markovian. Interestingly, these considerations have been taken into account in a previous study on \textit{Aedes} mosquitoes population dynamics, where simulations were employed to accommodate non-Markovian transitions \cite{focks1993dynamic}. 

While most traditional epidemiological compartment models assume exponentially distributed sojourn times, there exist studies that have considered relaxing this assumption to offer a more precise representation of transmission dynamics \cite{feng2021mosquito,nowzari2015general,forien2021estimating}. One effective approach is to approximate the sojourn time distribution with an Erlang distribution \cite{vajdi2023non,feng2021mosquito}. This is because a random variable with an Erlang distribution can be seen as the sum of multiple exponentially distributed random variables.  Consequently, by partitioning an epidemiological state into several phases characterized by exponential transition times, traditional compartment models can provide an approximate description of the non-Markovian process. Indeed, rather than relying solely on Erlang distributions, an even more accurate approximation can be achieved by utilizing a broader class of distributions known as phase-type distributions \cite{nowzari2015general}.

In this paper, we present a novel mechanistic model that captures the intricate relationships between vector populations and environmental variables to more accurately describe dengue transmission, allowing for the development of a more accurate global dengue risk map. Our model leverages the temperature-dependent development rates of mosquitoes more accurately compared to existing models. Drawing inspiration from a recent study \cite{feng2021mosquito} on non-Markovian epidemiological models, we initially formulate integral equations for the non-Markovian dynamics of the mosquito life cycle. Subsequently, we illustrate how assuming a phase-type distribution for transitions simplifies the integral equations into ordinary differential equations (ODEs). While this model is specifically designed for dengue transmission, our approach can be extended to other relevant epidemiological systems.

One of the primary objectives of this paper is to develop a global, time-dependent risk metric that quantifies the environmental suitability for dengue outbreaks. Here, we utilize available epidemic data from various locations around the world to create this risk map. Interestingly, our map designates a high risk to areas that are not identified as high risk in other dengue risk maps. To accomplish this, we fit the outbreak data to the dengue transmission model using a particle filter algorithm, enabling the estimation of the environment's carrying capacity and mosquito biting rates. Subsequently, we develop a numerical risk model that incorporates these estimated values, along with daily average temperature and precipitation, to calculate the global spatiotemporal dengue outbreak risk based on the dengue reproductive number and vector population. Implementation of the risk model is available online   \cite{picturee,yi2023picturee}

The main contributions of this study are: a. Developing a new model for the mosquito life cycle that accounts for temperature-dependent development rates. This model improves upon other existing models, which often provide only approximate descriptions of the complex non-Markovian process; b. Creating a global dengue risk map with daily temporal resolution and a spatial resolution of 0.25 degrees in longitude and latitude. The structure of the remaining sections of this paper are as follows: Section 2 presents the integral equations used to describe the mosquito life cycle, emphasizing the non-Markovian nature of transitions between different life stages. Subsequently, we demonstrate the reduction of these integral equations to corresponding ordinary differential equations (ODEs). Moving forward, in Section 3, we incorporate dengue virus transmission into the mosquito life cycle model, resulting in the virus transmission model. In Section 4, we outline the methodology employed for developing an environment carrying capacity model based on precipitation and available outbreak data. Furthermore, we detail the construction of our risk model, which is based on the transmission model and the estimated parameters. Lastly, Section 5 concludes the paper by summarizing the contributions of our novel mechanistic model and discussing potential avenues for future research
\section{mosquitoes life cycle model}
\begin{figure*}[t]
\centering
   \includegraphics[width=1\columnwidth]{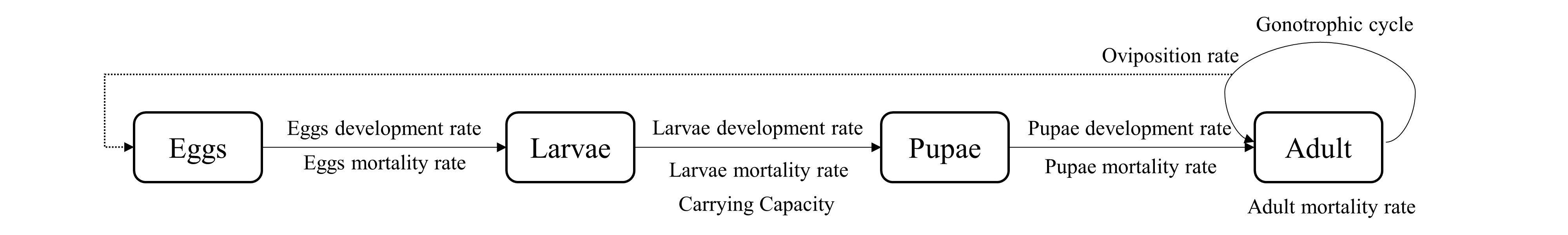} %
   \caption{The \textit{Aedes} mosquitoes life cycle. The rates depicted in the figure are temperature-dependent, which consequently change over time.}  
\label{lifecycle}
\end{figure*}

To determine the population of mosquitoes at various stages of growth, we use a model with four states representing the egg, larva, pupa, and adult stages. Figure \ref{lifecycle} illustrates the different states and the corresponding rates in the model. It is well established that the development time for each life-cycle stage is temperature dependent and the transition to the next stage occurs after the development is completed \cite{focks1993dynamic,eisen2014impact,couret2014temperature,yang2011follow,yang2009assessing2,
rossi2014modelling,farnesi2009embryonic,mordecai2017detecting}. Therefore, calculating the mosquito population requires tracking the extent of development across the life-cycle stages. We adopt the following master equations to determine the total population in each state of the life cycle.      
\begin{equation}
\begin{split}
& l(t)=\int_{0}^{t} \lambda_{e,l}(u)\ \mathcal{P}_{l,p}(u,t)\ \mathcal{P}_{l,d}(u,t)\ du+l(0)\ \mathcal{P}_{l,p}(0,t)\ \mathcal{P}_{l,d}(0,t)\\
& p(t)=\int_{0}^{t} \lambda_{l,p}(u)\ \mathcal{P}_{p,a}(u,t)\ \mathcal{P}_{p,d}(u,t)\ du+p(0)\ \mathcal{P}_{p,a}(0,t)\ \mathcal{P}_{p,d}(0,t)\\
& a(t)=\int_{0}^{t} \lambda_{a}(u)\ \mathcal{P}_{a,e}(u,t)\ \mathcal{P}_{a,d}(u,t)\ du+a(0)\ \mathcal{P}_{a,e}(0,t)\ \mathcal{P}_{a,d}(0,t)\\
& e(t)=\int_{0}^{t} \lambda_{e}(u)\ \mathcal{P}_{e,l}(u,t)\ \mathcal{P}_{e,d}(u,t)\ du+e(0)\ \mathcal{P}_{e,l}(0,t)\ \mathcal{P}_{e,d}(0,t)
\end{split}
\label{inteq}
\end{equation}

\begin{figure*}[t]
\centering
   \includegraphics[width=0.5\columnwidth]{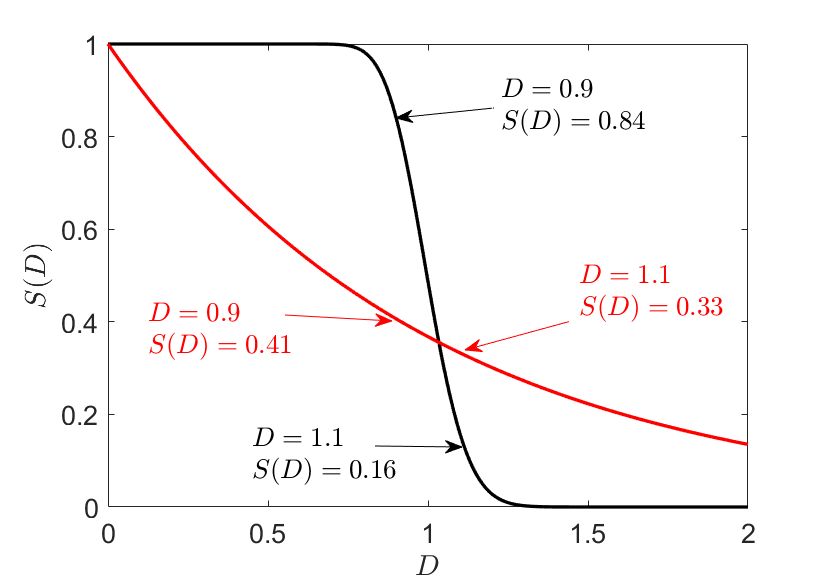} %
   \caption{The survival function of the Erlang distribution is represented by the black curve, while that of the exponential distribution is depicted by the red curve.. Both distributions have a mean equal to one.}  
\label{surv}
\end{figure*}

Here, we define $l(t)$, $p(t)$, $a(t)$, and $e(t)$ as the quantities representing the number of larvae, pupae, adult female mosquitoes, and eggs, respectively, irrespective of their developmental stage. The term $\lambda_{e,l}(u)$ represents the rate of larval emergence at time $u$. In other words, $\lambda_{e,l}(u) du$ corresponds to the number of eggs that hatch into larvae during the time interval $[u, u+du]$. Furthermore, $\mathcal{P}_{l,p}(u,t)$ denotes the probability that a larva, which emerged at time $u$, remains in the larval stage and has not yet transformed into a pupa by time $t$, given that it has survived up to time $t$. This probability is unity unless the larva completes its development, at which point it becomes zero. Assuming the larval development rate at temperature $T$ is denoted as $\gamma_{l,p}(T)$, we can express the extent of development from time $u$ to $t$ as 
\begin{equation}\label{dum1}
 \mathcal{D}_{l,p}(u,t)=\int_{u}^{t} \gamma_{l,p}(T(\tau))\ d\tau,
\end{equation} 
where $\mathcal{D}_{l,p}=1$ indicates full development. This equation aligns with an approximation used to calculate the duration of the development period when the environmental temperature is variable [reference]. With the development extent quantified as we have described, we can formulate the probability of the larval stage continuing from time $u$ to time $t$ as 
\begin{equation}
\mathcal{P}_{l,p}(u,t)=\mathcal{S}_{l,p}(\mathcal{D}_{l,p}(u,t)),
\end{equation}
where $\mathcal{S}_{l,p}$ represents the survival function of a probability density function $f(x)$ centered around one, i.e.,
\[
\mathcal{S}_{l,p}(\mathcal{D})=1-\int_0^D f(x) \ dx.
\]
Figure \ref{surv} shows the survival function for two different choices of density functions. 
The black curve is survival function for the gamma distribution with shape parameter $100$ and rate $0.01$. Considering this survival function as $\mathcal{S}_{l,p}(\mathcal{D})$, implies the probability of transitioning to the pupa stage when the larvae development is less than $0.9$ is smaller than $0.14$, and this probability increases to $0.84$ when the development reaches $1.1$. This is in agreement with the simulation method in \cite{focks1993dynamic}.  In contrast, utilizing the survival function of the exponential distribution with a rate of $1$, depicted by the red curve in the figure, results in $59$ percent of larvae progressing to the pupal stage before their development reaches $0.9$. In fact, we will demonstrate that describing the mosquitoes' life cycle using traditional ODEs with temperature-dependent rates is equivalent of employing the survival function of exponential distribution in the equations \ref{inteq}. Thus, for a more accurate representation of the life cycle, we utilize the integral equations \ref{inteq} with survival functions similar to the black curve shown in Figure \ref{surv}. Similar to the development rate, the mortality rate of larvae, $\gamma_{l,d}$, is temperature dependent. Consequently, we express the survival probability of the larvae, $\mathcal{P}_{l,d}(u,t)$, as   
\begin{equation}\label{dum2}
 \mathcal{P}_{l,d}(u,t)=\mathcal{S}_{l,d}(\mathcal{A}_{l,p}(u,t)), \  \mathcal{A}_{l,d}(u,t)=\int_{u}^{t} \gamma_{l,d}(T(\tau))\ d\tau.
\end{equation}
To maintain the consistency in the presentation of equations \ref{inteq}, we write the larvae survival probability in terms of a general survival function $\mathcal{S}_{l,d}$, but in our calculations, we use the exponential distribution survival function, which is $\mathcal{S}(x)=e^{-x}$. This choice of exponential distribution implies the assumption that the mortality of larvae is a non-homogeneous markovian process with a temperature dependent mortality rate.  

The master equations \ref{inteq} include the emergence rates as well as survival probabilities for the pupa, adult, and egg stages, which are defined in a manner similar to the probabilities and rates for the larva stage. To complete the life cycle description we need to express the emergence rates of each stage in terms of the parameters of the preceding stage. To achieve this, we analyze the derivative of $l(t)$ with respect to time to determine the pupal emergence rate. It is straight forward to show 
\begin{equation}
\begin{split}
 \frac{dl(t)}{dt}=&\lambda_{e,l}(t)\ \mathcal{P}_{l,p}(t,t)\ \mathcal{P}_{l,d}(t,t)
\\&+\int_{0}^{t} \lambda_{e,l}(u)\ \frac{ \partial \mathcal{P}_{l,p}(u,t)}{\partial t}\ \mathcal{P}_{l,d}(u,t)\ du-l(0)\ \frac{ \partial \mathcal{P}_{l,p}(0,t)}{\partial t}\ \mathcal{P}_{l,d}(0,t)
\\&+\int_{0}^{t} \lambda_{e,l}(u)\ \mathcal{P}_{l,p}(u,t)\ \frac{ \partial \mathcal{P}_{l,d}(u,t)}{\partial t}\ du-l(0)\ \mathcal{P}_{l,p}(0,t)\ \frac{ \partial \mathcal{P}_{l,d}(0,t)}{\partial t}.
\end{split}
\label{lamp}
\end{equation} 
The equation above demonstrates that the larvae rate of change consists of three components: the first line on the right-hand side of the equation represents the influx from the egg stage, while the remaining two lines represent the outflow to the pupae and dead larvae stages, respectively.
Taking into account that $\mathcal{P}_{l,p}(t,t)=1$ and $\mathcal{P}_{l,d}(t,t)=1$, the inflow at time $t$ is $\lambda_{e,l}(t)$, which is consistent with our initial definition of $\lambda_{e,l}$. Additionally, $ -dt \partial \mathcal{P}_{l,p}(u,t)/\partial t$ is the probability that larvae that emerged at time $u$ transition to the pupal stage within the time interval $[t,t+dt]$, given their survival until time $t$.  Similarly, $ -\partial \mathcal{P}_{l,d}(u,t)/\partial t$ represents the mortality probability density of the larvae at time $t$. Indeed, if we assume $\mathcal{P}_{l,d}$ is written in terms of the survival function for the exponential distribution we obtain  $ -\partial \mathcal{P}_{l,d}(u,t)/\partial t= \mathcal{P}_{l,d}(u,t) \gamma_{l,d}(T(t))$. 

Thus, our analysis of the terms in the expression for $dl(t)/dt$ demonstrates that the second line in equation \ref{lamp} represents the emergence rate of pupae at time $t$. With a similar reasoning, the emergence rates in the equations \ref{inteq} can be expressed as follows
\begin{equation}
\begin{split}
& \lambda_{l,p}(\tau)=-\int_{0}^{\tau} \lambda_{e,l}(u)\ \ \frac{ \partial \mathcal{P}_{l,p}(u,\tau)}{\partial \tau}\ \mathcal{P}_{l,d}(u,\tau)\ du-l(0)\ \frac{ \partial \mathcal{P}_{l,p}(0,\tau)}{\partial \tau}\ \mathcal{P}_{l,d}(0,\tau)\\
& \lambda_{a}(\tau)=\frac{1}{2}\lambda_{p,a}(\tau)+\lambda_{a,a}(\tau)\\
&\lambda_{p,a}(\tau)=-\int_{0}^{\tau} \lambda_{l,p}(u)\ \frac{ \partial\mathcal{P}_{p,a}(u,\tau)}{\partial \tau}\ \mathcal{P}_{p,d}(u,\tau)\ du-p(0)\ \frac{ \partial\mathcal{P}_{p,a}(0,\tau)}{\partial \tau}\ \mathcal{P}_{p,d}(0,\tau)\\
&\lambda_{a,a}(\tau)= -\int_{0}^{\tau} \lambda_{a}(\tau)\ \frac{\partial\mathcal{P}_{a,e}(u,\tau)}{\partial\tau}\ \mathcal{P}_{a,d}(u,\tau)\ du-a(0)\ \frac{\partial\mathcal{P}_{a,e}(0,\tau)}{\partial\tau}\ \mathcal{P}_{a,d}(0,\tau)\\
&\lambda_{e}(\tau)=ov(T(\tau))\lambda_{a,a}(\tau)\\
& \overline{\lambda}_{e,l}(\tau)=-\int_{0}^{\tau} \lambda_{e}(u)\ \frac{\partial\mathcal{P}_{e,l}(u,\tau)}{\partial \tau}\ \mathcal{P}_{e,d}(u,\tau)\ du-e(0)\ \frac{\partial\mathcal{P}_{e,l}(0,\tau)}{\partial \tau}\ \mathcal{P}_{e,d}(0,\tau)\\
&\lambda_{e,l}(\tau)=\left(1-\frac{l(\tau)}{C(\tau)}\right)\overline{\lambda}_{e,l}(\tau)
\end{split}
\label{emrate}
\end{equation}
We can see that the adult mosquitoes emergence rate, $\lambda_a$, is the sum of $1/2\lambda_{p,a}$, which is the inflow from pupa state to adult female state, and $\lambda_{a,a}$ that represents the number adult female mosquitoes completed a gonotrophic cycle and oviposition and starting a new gonotrophic cycle. Here, we assume that half of the pupae population are female, and the other half are male. This two terms are illustrated in figure \ref{lifecycle}. Additionally, we assume that the number of eggs laid within a time interval $[\tau, \tau+d\tau]$, denoted by $d\tau\lambda_e(\tau)$, is the product of two factors: the number of adult mosquitoes that have completed the gonotrophic cycle within that interval, $d\tau\lambda_{a,a}(\tau)$, and the average number of eggs laid per adult female mosquito at the temperature $T(\tau)$, denoted as $ov(T(\tau))$. We also account for the environment's carrying capacity, denoted as $C(\tau)$, by constraining the number of eggs that can hatch into the larval state. This constraint is reflected in the expression for $\lambda_{e,l}$.
\subsection{Describing the mosquitoes life cycle using ODEs }\label{odesec}
Here we show how to reduce the life cycle integral equations \ref{inteq} to ODEs when the probabilities $\mathcal{P}(u,t)$ in these equations are given in terms of the survival function of phase-type distributions. A phase-type distribution is the probability distribution for the absorption time in a continuous-time Markov chain with one absorbing state\cite{cox1977theory,nowzari2015general}. It is known that any positive-valued distribution can be approximated using a phase-type distribution. For instance, a random variable $X$ that follows an Erlang distribution can be exactly represented as a phase-type distribution. In this case, $X$ can be perceived as the time taken to reach an absorbing state when a continuous-time Markov process with a number of successive transient states and one final absorbing state is initiated from the first state. Formally, a phase-type distribution with $m$ transient states is defined by a pair $(\bm{\alpha},\bm{\mathcal{Q}})$. The $1\times m$ row vector $\bm{\alpha}$ is the initial distribution of the Markov chain across the $m$ transient states and $(\bm{\mathcal{Q}})$ is an $m \times m$ matrix of transition rates among these transient states. Therefore, the generator matrix $\bm{G}$ for the Markov chain is given by
\[
\bm{G} = 
\begin{pmatrix}
 \bm{\mathcal{Q}}_{m\times m}&\bm{q}_{m\times 1}\\
 \bm{0}_{1\times m}&0
\end{pmatrix}, \  \bm{q}= -\bm{\mathcal{Q}} \ \bm{1},
\]
where $\bm{1}$ is a column vector with all entries set to 1.
The cumulative distribution, $\mathcal{C}$, and survival function, $\mathcal{S}$, for a phase-type distribution are
\[
\mathcal{C}(t)=1-\bm{\alpha}\ e^{t\bm{\mathcal{Q}}}\ \mathbf{1},\ \mathcal{S}(t)=\bm{\alpha}\ e^{t\bm{\mathcal{Q}}}\ \mathbf{1}
\]
To clarify the procedure of converting the integral equations \ref{inteq} into ODEs, we will transform the integral equation describing the larvae population into an equivalent set of differential equations. Assuming  $\mathcal{P}_{l,p}(u,t)$ and $\mathcal{P}_{l,d}(u,t)$ are defined in terms of phase-type distribution survival functions,
\begin{equation}
\begin{split}
&\mathcal{P}_{l,p}(u,t)=\mathcal{S}_{l,p}(\mathcal{D}_{l,p})=\bm{\alpha}_{l,p}\ e^{\mathcal{D}_{l,p}\bm{\mathcal{Q}}_{l,p}}\ \mathbf{1},\\
& \mathcal{P}_{l,d}(u,t)=\mathcal{S}_{l,d}(\mathcal{A}_{l,d})=\bm{\alpha}_{l,d}\ e^{\mathcal{A}_{l,d}\bm{\mathcal{Q}}_{l,d}}\ \mathbf{1},
\end{split}
\end{equation}
we can write $l(t)$ as the outcome of the matrix multiplication between a row vector $\bm{l}$ and an all-ones column vector $\bm{1}$ as follows,
\begin{equation}
\begin{split}
 l(t)&=\bm{l}(t)\ \bm{1},\\
\bm{l}(t)&=\int_{0}^{t} \lambda_{e,l}(u) \left( \bm{\alpha}_{l,p}\ e^{\mathcal{D}_{l,p}(u,t)\bm{\mathcal{Q}}_{l,p}}\right)\otimes\left( \bm{\alpha}_{l,d}\ e^{\mathcal{A}_{l,d}(u,t)\bm{\mathcal{Q}}_{l,d}}\right)\ du+l(0)\left( \bm{\alpha}_{l,p}\ e^{\mathcal{D}_{l,p}(0,t)\bm{\mathcal{Q}}_{l,p}}\ \right)\otimes\left( \bm{\alpha}_{l,d}\ e^{\mathcal{A}_{l,d}(0,t)\bm{\mathcal{Q}}_{l,d}}\right).
\end{split}
\end{equation}
We have expressed $\bm{l}(t)$ using the Kronecker product of row vectors, denoted by the symbol $\otimes$ in the equation above. Given the definitions of $\mathcal{D}_{l,d}(u,t)$ and $\mathcal{A}_{l,d}(u,t)$ in equations \ref{dum1} and \ref{dum2} respectively, it is straight forward to calculate the derivative of $\bm{l}(t)$ with respect to time, 
\begin{equation}\label{lode}
 \frac{d\bm{l}(t)}{dt}=\lambda_{e,l}(t)\ \bm{\alpha}_{l,p} \otimes \bm{\alpha}_{l,d}+\bm{l}(t)\left( \gamma_{l,p}(t) \bm{\mathcal{Q}}_{l,p}\otimes \bm{I}_{l,d}  + \gamma_{l,d}(t)  \bm{I}_{l,p}\otimes\bm{\mathcal{Q}}_{l,d} \right)
\end{equation}
Where $\bm{I}_{l,p}$ and $\bm{I}_{l,d}$ are identity matrices that have dimensions that match those of $\bm{\mathcal{Q}}_{l,p}$ and $\bm{\mathcal{Q}}_{l,d}$, respectively. In calculating the derivative of $\bm{l}(t)$, the following equations were employed.

\begin{equation}
\begin{split}
&\frac{ \partial\  e^{\mathcal{D}_{l,p}(u,t)\bm{\mathcal{Q}}_{l,p}}}{\partial t}=\gamma_{l,p}(t)\ e^{\mathcal{D}_{l,p}(u,t)\bm{\mathcal{Q}}_{l,p}} \bm{\mathcal{Q}}_{l,p} \\
&\frac{ \partial\  e^{\mathcal{A}_{l,d}(u,t)\bm{\mathcal{Q}}_{l,d}}}{\partial t}=\gamma_{l,d}(t)\ e^{\mathcal{A}_{l,d}(u,t)\bm{\mathcal{Q}}_{l,d}} \bm{\mathcal{Q}}_{l,d}
\end{split}
\end{equation}

Differential equation \ref{lode} describes the temporal evolution of the vector $\bm{l(t)}$, and the larvae population is the sum of the elements in $\bm{l(t)}$. Similar procedure can be used to convert the integral equations that describe the pupa, egg and adult mosquito populations into differential equations governing the corresponding vectors  $\bm{p(t)}$, $\bm{e(t)}$, $\bm{a(t)}$. Furthermore, the emergence rates, which are given in equations \ref{emrate}, could be expressed in terms of the components of the vectors $\bm{l(t)}$, $\bm{p(t)}$, $\bm{e(t)}$, $\bm{a(t)}$; for instance, 
\[
\lambda_{l,p}(t)=-\gamma_{l,p}(t)\ \bm{l}(t)\left( \bm{\mathcal{Q}}_{l,p}\otimes \bm{I}_{l,d}\right)\ \bm{1}=\gamma_{l,p}(t)\ \bm{l}(t)\ (\bm{q}_{l,p}\otimes\bm{1}),
\]
where $\bm{q}_{l,p}= -\bm{\mathcal{Q}}_{l,p} \ \bm{1}$.

\begin{figure*}[t]
\centering
   \includegraphics[width=0.75\columnwidth]{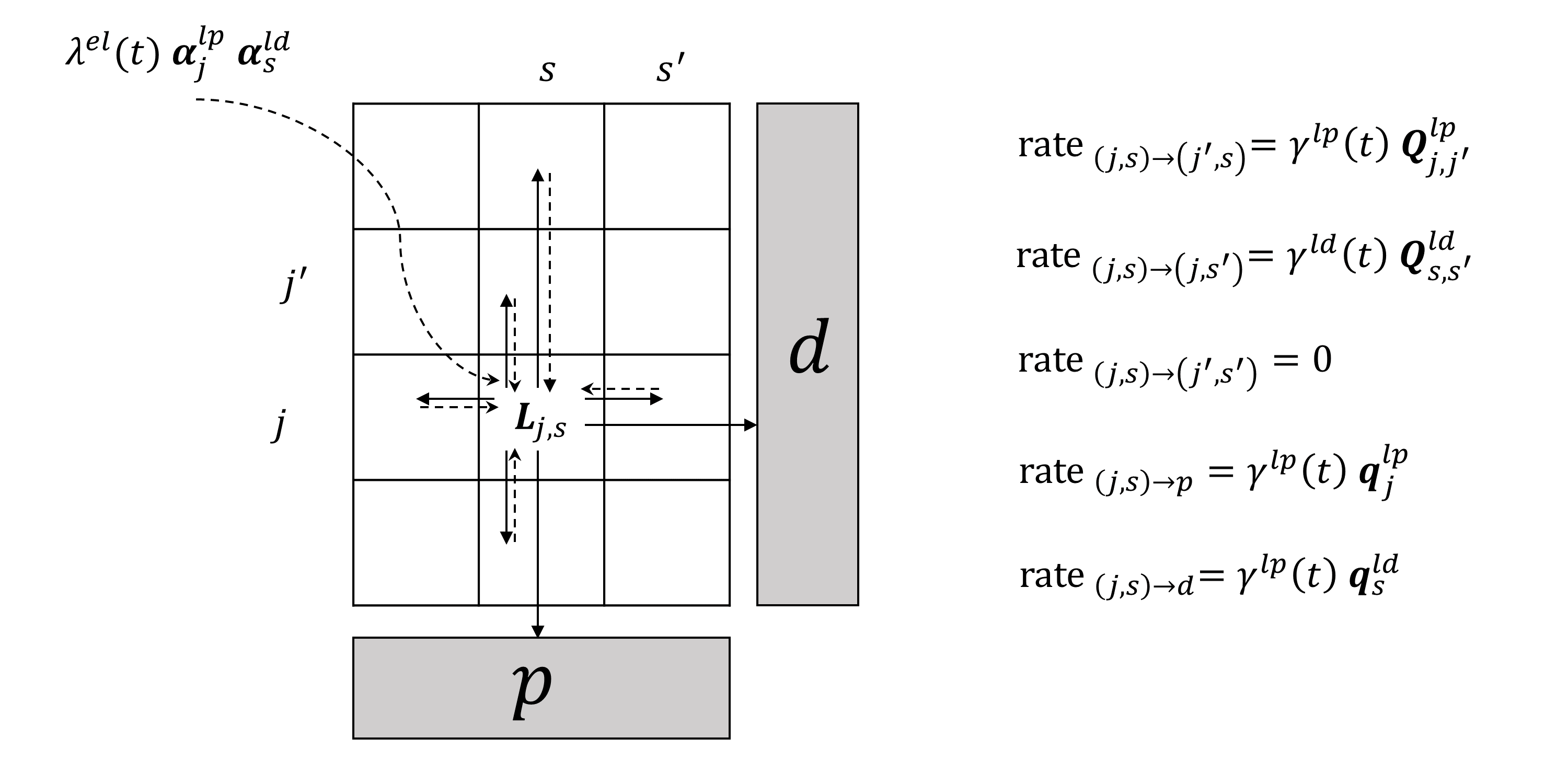} %
   \caption{The diagram illustrates Equation \ref{lmateq}, which describes the transition from the larva to pupa and dead larva states, represented as the absorbing states 'p' and 'd,' respectively. By utilizing phase-type distributions for the transition processes in the larva state, we effectively partition the larva state into several auxiliary compartments that represent the history of the original processes. The figure illustrates these larva compartments and the corresponding transition rates as given by Equation \ref{lmateq}.}  
\label{lmatf}
\end{figure*}

The components of vectors $\bm{l(t)}$, $\bm{p(t)}$, $\bm{e(t)}$ and $\bm{a(t)}$ can be rearranged to provide a more intuitive picture of the ODEs describing the mosquito life-cycle. To this end, assume  that $\bm{\mathcal{Q}}_{l,p}$ and $\bm{\mathcal{Q}}_{l,d}$ in equation \ref{lode}, are $m_1\times m_1$ and $m_2\times m_2$ matrices, respectively, indicating that the corresponding phase-type distributions have $m_1$ and $m_2$ transient states. If we rearrange the row vector $\bm{l}(t)$ into a $m_1\times m_2$ matrix $\bm{L}(t)$ defined as:
\[
\bm{L}(t)_{j,s}=\bm{l}(t)_{(j-1)\times m_{2}+s},\   j\in\{1,\cdots,m_1\}, \ s\in\{1,\cdots,m_2\},
\]
equation \ref{lode} can be rewritten as 
 \begin{equation}\label{lmateq}
\frac{d\bm{L}_{j,s}}{dt}=\lambda_{e,l}(t)\ (\bm{\alpha}_{l,p})_j \  (\bm{\alpha}_{l,d})_s+\gamma_{l,p}(t)\sum_{k=1}^{r_1}(\bm{\mathcal{Q}}_{l,p})_{k,j}\bm{L}_{k,s}+\gamma_{l,d}(t)\sum_{k=1}^{r_2}(\bm{\mathcal{Q}}_{l,d})_{k,s}\bm{L}_{j,k},
\end{equation}
where the larvae population can be expressed as $l(t)=\sum_{j,s}^{m_{1},m_{2}}\bm{L}(t)_{j,s}$. The definition of $\bm{L}$ implies that the total larvae population is distributed among $m_1 \times m_2$ compartments and the elements of $\bm{L}$ represent the compartments' population. Additionally, non-homogeneous Markovian transitions among these compartments are possible and they are governed by the $\bm{\mathcal{Q}}$ matrices. Figure \ref{lmatf} illustrates the larvae compartments and the transition rates given by equation \ref{lmateq}. The pupa and dead larva states are depicted as the absorbing states "p" and "d" in the figure. It is straight forward to show that the pupa emergence rate is 
\[
\lambda_{l,p}(t)=\gamma_{l,p}(\tau)\sum_{j,s}^{m_{1},m_{2}}\bm{L}(t)_{j,s}\ (\bm{q}_{l,p})_j.
\]

Similar interpretation of the ODEs applies to the other states in the mosquito life-cycle, meaning that, each state divides into several sub-states representing different levels of developments.  

The simplest phase-type distribution consists of only one state, and the initial distribution vector and the transition rate matrix are 
\[
\bm{\mathcal{Q}}=\left(-1\right),\  \bm{\alpha}=\left(1\right).
\]
This trivial phase-type distribution is an exponential distribution with mean one. If we assume all the probability functions $\mathcal{P}$ in the integral equations \ref{inteq} are expressed in terms of the survival function of this particular phase-type distribution, then the corresponding system of ODEs becomes,  

\begin{equation}\label{odem}
\begin{split}
&\frac{dl(t)}{dt}=\left(1-\frac{l(t)}{C(t)}\right) \gamma_{e,l}(t) e(t)\  -\gamma_{l,p}(t) \ l(t)-\gamma_{l,d}(t) \ l(t)\\
&\frac{dp(t)}{dt}=\gamma_{l,p}(t)\ l(t)\  -\gamma_{p,a}(t) \ p(t)-\gamma_{p,d}(t) \ p(t)\\
&\frac{da(t)}{dt}=\frac{1}{2}\gamma_{p,a}(t)\ p(t)-\gamma_{a,d}(t) \ a(t)\\
&\frac{de(t)}{dt}=ov(t)\ \gamma_{a,e}(t)\ a(t)\  -\gamma_{e,l}(t) \ e(t)-\gamma_{e,d}(t) \ e(t)\\
\end{split}
\end{equation}
The system of ODEs above is often used in mechanistic models to calculate mosquito populations. However, these equations allow transitions from a life-cycle stage even if the development in that stage is not complete. This is due to the form of exponential distribution survival function  which is depicted in figure \ref{surv}. A better choice of survival functions that prevent immature transitions is portrayed in the same figure by the black curve which is the Erlang distribution survival function. The Erlang distribution with shape parameter $J$ and mean one can be viewed as a phase-type distribution with $J$ transient states. The initial distribution vector and the transition rate matrix that are as follows 
\begin{equation}\label{delmat}
\begin{split}
\bm{\alpha}=\begin{pmatrix}
1 & 0&0&\cdots & 0&0 
\end{pmatrix}_{1\times J},\ 
\bm{\mathcal{Q}}=\begin{pmatrix}
-J & J&0&\cdots & 0&0 \\
0&-J & J&\cdots & 0&0 \\
\vdots  & \vdots & \vdots &\vdots  & \vdots & \vdots  \\
0  & 0 & 0 &\cdots  & -J & J  \\
0  & 0 & 0 &\cdots  & 0 & -J  \\
\end{pmatrix}_{J\times J}
\end{split}
\end{equation}

  \begin{figure*}[t]
\centering
   \includegraphics[width=0.85\columnwidth]{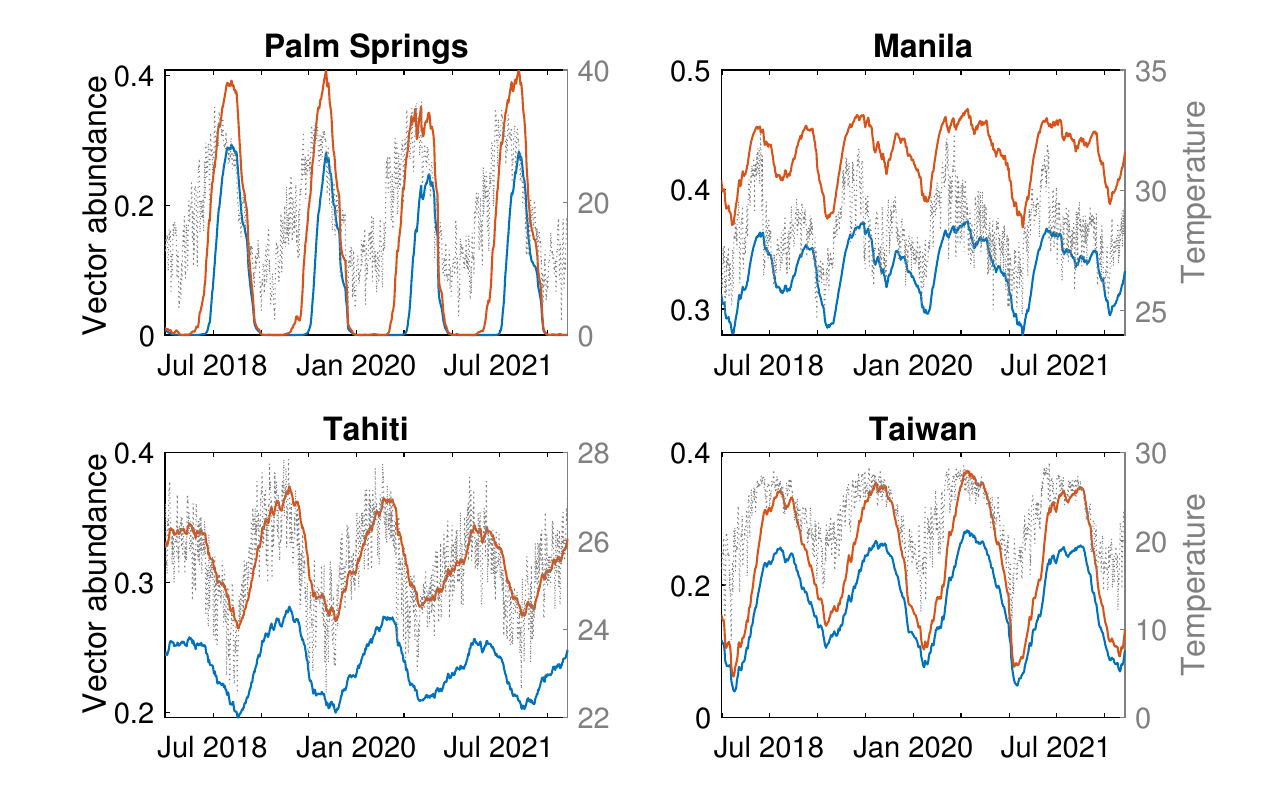} %
   \caption{The figures show simulated vector abundance (arbitrary units) using a non-Markovian model (blue curves) and a Markovian model (red curves). In these simulations, we utilize the daily average temperature for each location indicated in the figure, and the temperature profiles are depicted using gray curves.}  
\label{compjs}
\end{figure*}
Assuming that the development probabilities $\mathcal{P}_{l,p}$, $\mathcal{P}_{p,a}$, $\mathcal{P}_{a,e}$ and $\mathcal{P}_{e,l}$ in the integral equations \ref{inteq} are defined using the survival function of an Erlang distribution with mean one and shape parameter $J$, and that the mortality probabilities $\mathcal{P}_{.,d}$ are expressed in terms of the survival function an exponential distribution with mean one, the resulting system of ODEs is given by 

\begin{equation}\label{odeer}
\begin{split}
&\frac{1}{J}\frac{d\bm{L}_{j}}{dt}=\left(1-\frac{l(t)}{C(t)}\right)\gamma_{e,l}(t)\bm{E}_{J}\ 1_{j=1}+\gamma_{l,p}(t)\left(1_{j>1}\ \bm{L}_{j-1}-\bm{L}_{j}\right)-\frac{1}{J}\gamma_{l,d}(t)\bm{L}_{j} \\
&\frac{1}{J}\frac{d\bm{P}_{j}}{dt}=\gamma_{l,p}(t)\bm{L}_{J}\ 1_{j=1}+\gamma_{p,a}(t)\left(1_{j>1}\ \bm{P}_{j-1}-\bm{P}_{j}\right)-\frac{1}{J}\gamma_{p,d}(t)\bm{P}_{j} \\
&\frac{1}{J}\frac{d\bm{A}_{j}}{dt}=\left(\frac{1}{2}\gamma_{p,a}(t)\bm{P}_{J}+\gamma_{a,e}(t)\bm{A}_{J}\right)\ 1_{j=1}+\gamma_{a,e}(t)\left(1_{j>1}\ \bm{A}_{j-1}-\bm{A}_{j}\right)-\frac{1}{J}\gamma_{a,d}(t)\bm{A}_{j} \\
&\frac{1}{J}\frac{d\bm{E}_{j}}{dt}=ov(t)\gamma_{a,e}(t)\bm{A}_{J}\ 1_{j=1}+\gamma_{e,l}(t)\left(1_{j>1}\ \bm{E}_{j-1}-\bm{E}_{j}\right)-\frac{1}{J}\gamma_{e,d}(t)\bm{E}_{j} \\
\end{split}
\end{equation}
where $j\in\{1,\cdots,J\}$, and $1_{j=1}$ is an indicator function. This function is equal to zero for all values of $j$ except when $j$ is equal to one, in which case it takes the value of one. Employing the phase-type distribution with $J$ transient states implies the assumption that the mosquito population in each life-cycle stage is distributed across $J$ sub-states representing different levels of development. Thus the populations of different life-cycle stages are written as  
\begin{equation}
\begin{split}
&l(t)=\sum_{j=1}^{J}\bm{L}(t)_{j},\ p(t)=\sum_{j=1}^{J}\bm{P}(t)_{j},\ a(t)=\sum_{j=1}^{J}\bm{A}(t)_{j},\ e(t)=\sum_{j=1}^{J}\bm{E}(t)_{j} 
\end{split}
\end{equation}
\begin{figure*}[t]
\centering
   \includegraphics[width=0.85\columnwidth]{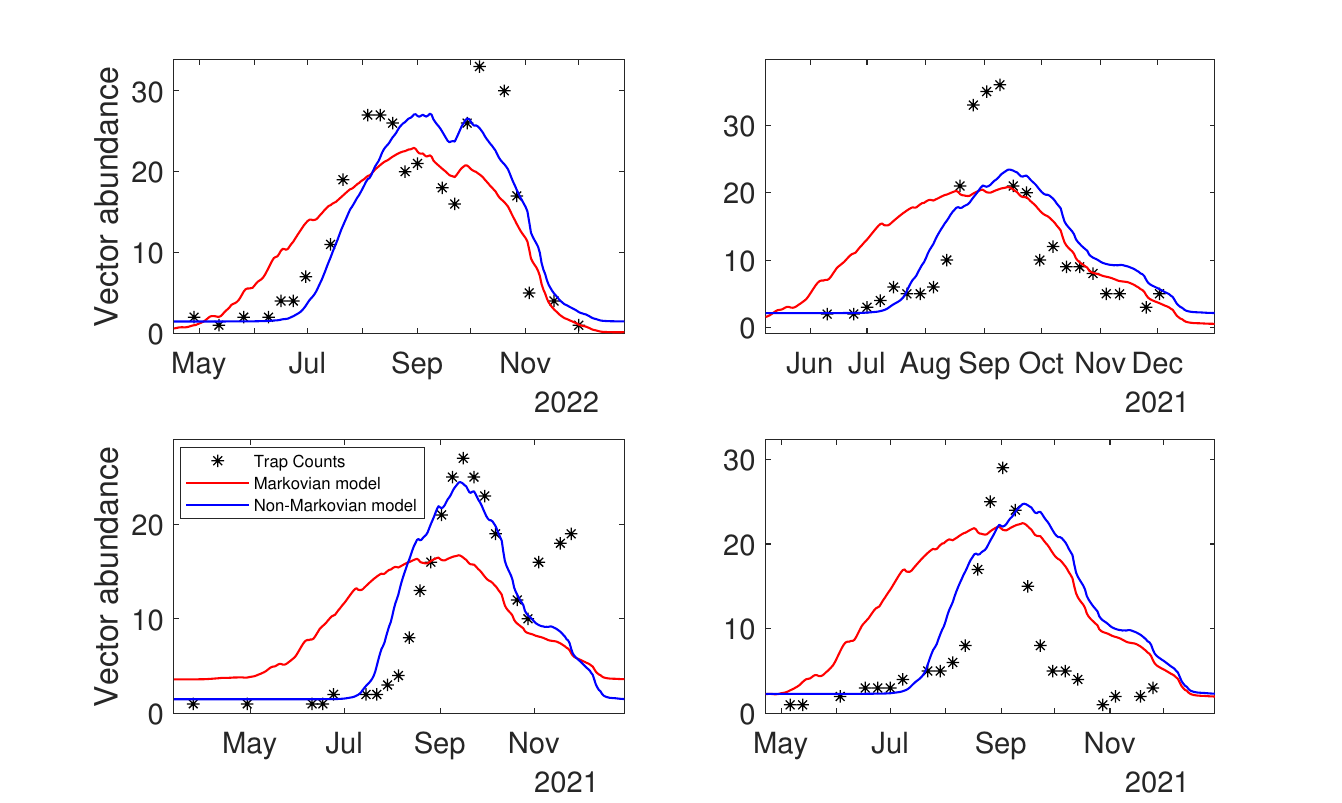} %
   \caption{Comparing the \textit{Aedes aegypti} trap counts in four different locations in Coachella Valley, California, with simulated vector abundance using a non-Markovian model (blue curves) and a Markovian model (red curves).}  
\label{compps}
\end{figure*}

In order to demonstrate the impact of the distribution of development levels leading to a new life stage on the population of adult mosquitoes, we solved the equations \ref{odeer} for temperature profiles in four different locations twice: once with the assumption of $J=1$, which corresponds to exponential distribution, and once with the assumption of $J=100$, which only allows the transitions with the development levels around 100 percent. Figure \ref{compjs} shows the results of these calculations. We can observe using a value of $J=1$ for regions such as Manila and Tahiti, where the temperature varies between 23 and 32 degrees Celsius throughout a year, leads to a higher abundance of the adult mosquitoes compared to using a value of $J=100$. However, both curves have a similar profile over time, except that the curve for the value of $J=1$ is shifted upward. We can also note that in regions like Palm Springs, California, where the temperature stays below 15 degrees Celsius for an extended period each year, the difference between the curves for $J=1$ and $J=100$ is more distinct. Specifically, the difference between the maximum and minimum abundance on the curve corresponding to $J=100$ is smaller than that of $J=1$,  and for $J=1$, the maximum is reached at a later time compared to the case of $J=100$. 

In figure \ref{compps}, we have compared the time series of  mosquito counts obtained from trapping at four distinct sites in Palm Springs[] with the solutions of equations \ref{odeer}. The stars in the plots are the moving averages of the trap counts, calculated as $m_i=(c_{i-1}+c_{i}+c_{i+1})/3$, where $c_i$ is the trap counts. If $a^1_i$ represents the calculated adult mosquito population with $J=1$ and for the time of observation $i$, we assumed the observations at site $s$ have Poisson distributions with the means $k^1_sa^1_i+r^1_s$, where $r^1_s<2$. Using an MCMC method we find the values of $k^1_s$ and $r^1_s$ that best explain the observations for site s. The red curves in the plots show the fitted $k^1_sa^1_i+r^1_s$ curves. We also solved equations \ref{odeer} with $J=100$, and the blue curves depict these solutions fitted to observations. We note that for the traps shown in figure \ref{compps}, the model with $J=100$ is a better fit to the observed data compared to the model with $J=1$. This is due to the fact that the model with $J=100$ more accurately accounts for the longer development time of adult mosquito from egg during the cooler periods of the year. It is important to note that the trap counts within a specific area strongly depend on local factors such as available habitats, temperature condition within the habitats, and other factors that can impact the trapping process, such as wind pattern. Therefore,  fitting mechanistic models to the trap counts without a comprehensive understanding of these local factors may not provide a conclusive support for adopting a specific mechanistic model. 

Although the carrying capacity, development rates, and mortality rates in equations \ref{odeer} are time-dependent, it is constructive to examine the steady state of these equations assuming constant rates and constant carrying capacity. By setting the time derivatives on the left-hand-side of the equations to zero, we can find the steady state population of the eggs, larvae, pupae and adult mosquitoes. For the non-trivial equilibrium point, we have determined that the total larvae population is given by
\[
\bar{l}=C\left(1-\frac{1}{r_{0}}\right),
\]
where the basic offspring number $r_{0}$ can be computed using the following equation:
\begin{equation}
\begin{split}
r_{0}=\frac{ov}{\left(1+\frac{\gamma_{a,d}}{J\gamma_{a,e}}\right)^{J}-1}\left(1+\frac{\gamma_{e,d}}{J\gamma_{e,l}}\right)^{-J}\left(1+\frac{\gamma_{l,d}}{J\gamma_{l,p}}\right)^{-J}\left(1+\frac{\gamma_{p,d}}{J\gamma_{p,a}}\right)^{-J}. 
\end{split}\label{offspring}
\end{equation}
We have also found that the populations of eggs, pupae and adult mosquitoes are proportional to the larvae population. Therefore, a meaningful non-trivial equilibrium point can only exist if the value of $r_{0}$ is greater than 1, and we can show that this equilibrium point is stable for such values of $r_{0}$. In fact, the value of $r_0$ is the average number of viable adult mosquitoes produced by one adult mosquito. To understand the biological interpretation of $r_{0}$, note that the first term on the r.h.s of equation \ref{offspring} represents the expected number of eggs laid by one adult mosquito during its lifespan, and the subsequent terms provides the probability of an egg surviving to the adult form. As a result, the stability condition, $r_{0}>1$, implies that the mosquito population can only survive if each mosquito has more than one viable offspring on average. Furthermore, as $f(x)=(1+x/J)^J$ is an increasing function of $J$ for any positive $x$, we can deduce that the value of offspring number $r_0$ is maximum when $J=1$. This explains why models with $J>1$ result in lower mosquito abundance, as depicted in figure \ref{compjs}. 
\section{Dengue transmission model}\label{strm}
In the following, we modify the mosquito population model to describe dengue transmission between humans and  mosquitoes. Since the mosquito population may vary significantly during the year due to environmental factors such as temperature and habitat changes, any model that aims to describe the dynamics of dengue transmission must account for the mosquito population variation. To this end, we divide the adult female mosquitoes into susceptible, exposed and infectious populations and assume all newly emerged mosquitoes are susceptible and the infectious mosquitoes remain infectious until they die. During its gonotrophic cycle an infectious mosquito gives a number of bites to humans that can transmit the virus to a susceptible host. Similarly, a susceptible mosquito may acquire the virus through biting infectious humans. Since the extrinsic incubation period of dengue virus in mosquitoes depends on temperature, infected mosquitoes become infectious with a temperature dependent rate. In contrast, we assume that the intrinsic incubation and infectious periods of dengue in human do not significantly vary with temperature. Based on these considerations we can modify integral equations \ref{inteq} and add the human states in order to describe the transmission dynamics. However, we limit the probability functions in the integral equations to the Erlang distribution, as described in section \ref{odesec}. Therefore, we start from equations \ref{odeer} and incorporate the transmission dynamics by dividing the adult mosquito compartments $\bm{A}_j$ into susceptible, $\bm{A^s}_j$, exposed, $\bm{A^e}_j$ and infectious , $\bm{A^i}_j$. Furthermore, we can divide the $\bm{A^e}_j$ compartments into additional compartments to account for non-exponential transition from expose state to infectious state. But, we avoid these additional states to keep the presentation of equations simple. As a result, the third equations in set of equations \ref{odeer} is replaced by the following equations     
\begin{equation}\label{odeer2}
\begin{split}
&\frac{1}{J}\frac{d\bm{A^s}_{j}}{dt}=\left(\frac{1}{2}\gamma_{p,a}(t)\bm{P}_{J}+\gamma_{a,e}(t)\bm{A^s}_{J}\right)\ 1_{j=1}+\gamma_{a,e}(t)\left(1_{j>1}\ \bm{A^s}_{j-1}-\bm{A^s}_{j}\right)-\frac{1}{J}\gamma_{a,d}(t)\bm{A^s}_{j}-\lambda_{HV}(t)\bm{A^s}_{j} \\
&\frac{1}{J}\frac{d\bm{A^e}_{j}}{dt}=\gamma_{a,e}(t)\bm{A^e}_{J}\ 1_{j=1}+\gamma_{a,e}(t)\left(1_{j>1}\ \bm{A^e}_{j-1}-\bm{A^e}_{j}\right)-\frac{1}{J}\gamma_{a,d}(t)\bm{A^e}_{j}+\lambda_{HV}(t)\bm{A^s}_{j}-\gamma_{V}(t)\bm{A^e}_{j} \\
&\frac{1}{J}\frac{d\bm{A^i}_{j}}{dt}=\gamma_{a,e}(t)\bm{A^i}_{J}\ 1_{j=1}+\gamma_{a,e}(t)\left(1_{j>1}\ \bm{A^i}_{j-1}-\bm{A^i}_{j}\right)-\frac{1}{J}\gamma_{a,d}(t)\bm{A^i}_{j}+\gamma_{V}(t)\bm{A^e}_{j} 
\end{split}
\end{equation}
where $\lambda_{HV}$ is dengue transmission rate from humans to mosquitoes, and $\gamma_{V}^{-1}$ is the expected extrinsic period in mosquitoes. We note that the mosquitoes lays eggs regardless of their infection state. Therefore, the first term on the r.h.s of the equation describing the egg populations should be modified as follows   
\begin{equation}
\frac{1}{J}\frac{d\bm{E}_{j}}{dt}=ov(t)\gamma_{a,e}(t)\left(\bm{A^s}_{J}+\bm{A^e}_{J}+\bm{A^i}_{J}\right)\ 1_{j=1}+\gamma_{e,l}(t)\left(1_{j>1}\ \bm{E}_{j-1}-\bm{E}_{j}\right)-\frac{1}{J}\gamma_{e,d}(t)\bm{E}_{j} \\
\end{equation}

With regard to humans, we consider four sub-populations representing susceptible, $\bm{H^s}$, exposed, $\bm{H^e}$, infectious, $\bm{H^i}$, and recovered, $\bm{H^r}$, human populations. Assuming the transitions between the human states are Markovian we can write
\begin{equation}\label{odeer3}
\begin{split}
&\frac{d\bm{H^s}}{dt}=-\lambda_{VH}(t)\bm{H^s} \\
&\frac{d\bm{H^e}}{dt}=\lambda_{VH}(t)\bm{H^s}-\gamma_{H}\bm{H^e} \\
&\frac{d\bm{H^i}}{dt}=\gamma_{H}\bm{H^e}-\eta_{H} \bm{H^i}\\
&\frac{d\bm{H^r}}{dt}=\eta_{H} \bm{H^i}\\
\end{split}
\end{equation}
 In the equations above, $\lambda_{VH}$ is dengue transmission rate from mosquitoes to humans. Also, the constant parameters $\gamma_{H}^{-1}$ and $\eta_{H}^{-1}$ are, respectively, the expected intrinsic and infectious periods in humans. The infection transmission rate from vector to host can be expressed in terms of average number of bites a vector gives to the hosts in a gonotrophic cycle, $n_{B}$, and the expected length of gonotrophic cycle, $\gamma_{a,e}^{-1}$, as follows
 \begin{equation}
 \lambda_{VH}=n_B\gamma_{a,e}\phi_{VH}\frac{1}{N_H}\sum_j\bm{A^i}_j,
\end{equation}
where, $N_H$ is the human population and $\phi_{VH}$ represents the probability of an infected vector transmitting the infection to a host through a single bite. We have assumed that biting occurs at every stage of the gonotrophic cycle. But, if we restrict biting to only the later $k$ stages of the cycle, then in the equation above, the summation should be only on those stages.  Moreover, we should multiply the r.h.s of the equation by the ratio of $J$ (the total number of stages in the cycle) to $k$. If the probability of transmitting the infection from an infected human to a susceptible mosquito via a single bite is represented by $\phi_{HV}$, we can write the infection transmission rate from humans to vectors as
 \begin{equation}
 \lambda_{HV}=n_B\gamma_{a,e}\phi_{HV}\frac{1}{N_H}\bm{H^i}.
\end{equation}

\begin{figure*}[t]
\centering
   \includegraphics[width=1\columnwidth]{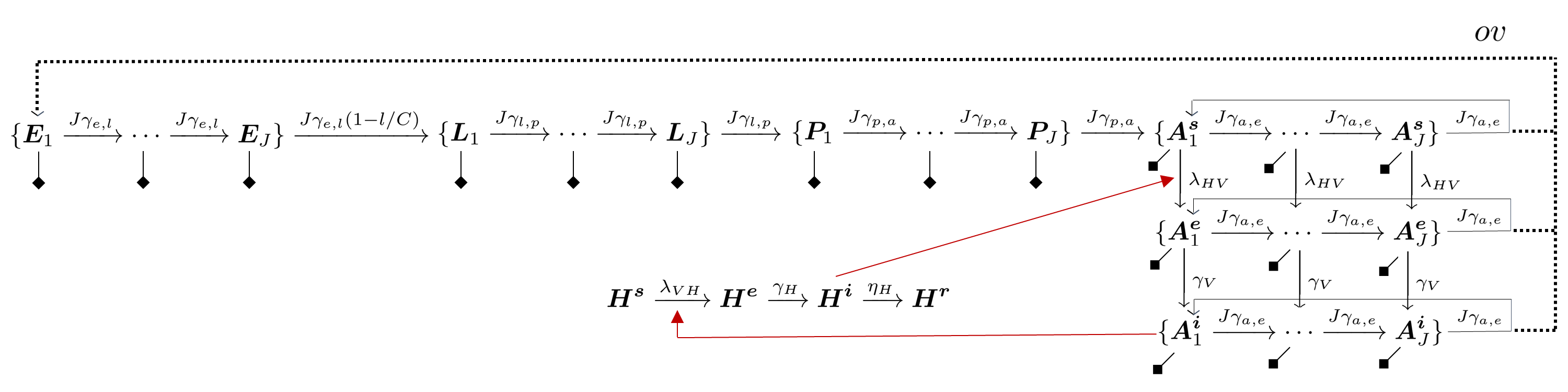} %
   \caption{The diagram depicts a non-Markovian dengue transmission model. The \textit{Aedes} mosquitoes life cycle, is modeled using the Erlang distribution with temperature-dependent transition rates. In the diagram, the populations of egg, larva, pupa, and adult mosquitoes are represented by $E_k$, $L_k$, $P_k$, and $A_k$, respectively, indicating different stages of development denoted by the subscript $k$. The human population is divided into susceptible individuals ($H^s$), exposed individuals ($H^e$), infectious individuals ($H^i$), and recovered individuals ($H^r$). Similarly, adult mosquitoes can be susceptible ($A^s$), exposed ($A^e$), or infectious ($A^i$). The red arrows indicate infectious transmission, while the black arrows represent transitions between two stages. Oviposition process is denoted by $ov$.}  
\label{tramod}
\end{figure*}
Figure \ref{tramod} illustrates the transmission model diagram. In case we consider that the mosquitoes' transition from the exposed state to the infectious state is non-Markovian, we need to divide $\bm{A^e}_j$ compartment into multiple states corresponding to different stages of the extrinsic incubation period. Therefore, the row that represents the mosquito exposed state in the diagram  
should be replicated. 
To find the dengue basic reproduction number, $R_0$,  under the assumption of constant temperature, we can analyze the stability of disease-free equilibrium.If we linearize the dengue transmission model around the steady state of mosquito population, using the techniques introduced in \cite{diekmann2010construction}, we can show that $R_0$ is 
\begin{equation}\label{repro}
R_0=\frac{\bar{a}}{N_H}n_B^2\gamma_{a,e}^2\phi_{HV}\phi_{VH} \gamma_{a,d}^{-1} \eta_H^{-1}\left(1+\frac{\gamma_{a,d}}{\gamma_V}\right)^{-1}.
\end{equation}  
The value of $R_0$ can be interpreted as the number female mosquitoes that become infected after introducing one infected female mosquitoes into a susceptible population of vectors and hosts. Although this interpretation is only valid for constant temperatures, it can still serve as an estimate for the degree of virus transmission if the temperature fluctuations are not too significant over a time frame longer than the inverse of the transition rates in the transmission model.
\section{numerical results}
Here, we apply the equations we developed in the previous section to construct a global risk map for dengue. This risk map is based on the entomological characteristics of \textit{Aedes aegypti} mosquito, as it is the main vector for dengue. To develop the risk map, we first estimate the environmental carrying capacity and biting rate for the mosquitoes. Using these estimated parameters and the other model parameters available in related literature, we can calculate the dengue reproduction number and the vector abundance. By correlating these variables with historical human dengue case data, we build a risk model that uses the reproduction number and vector abundance as predictive variables. This enables us to create a global spatio-temporal dengue risk map.   
\subsection{Estimating model parameters using particle filter}\label{carmod}
\begin{figure*}[t]
\centering
\subfloat[]{
   \includegraphics[width=0.45\columnwidth]{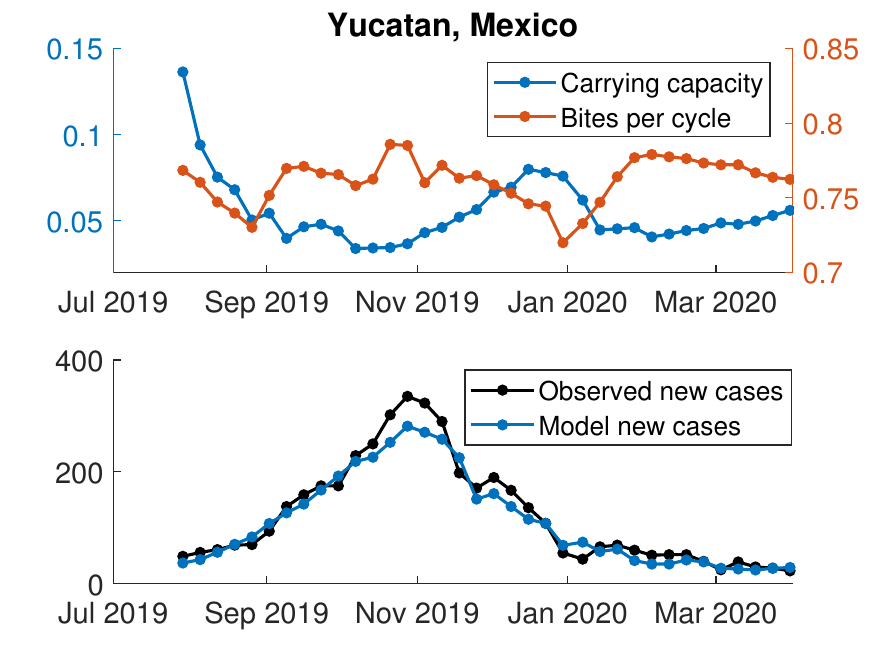} \label{fitlo}} %
   \subfloat[]{ \includegraphics[width=0.45\columnwidth]{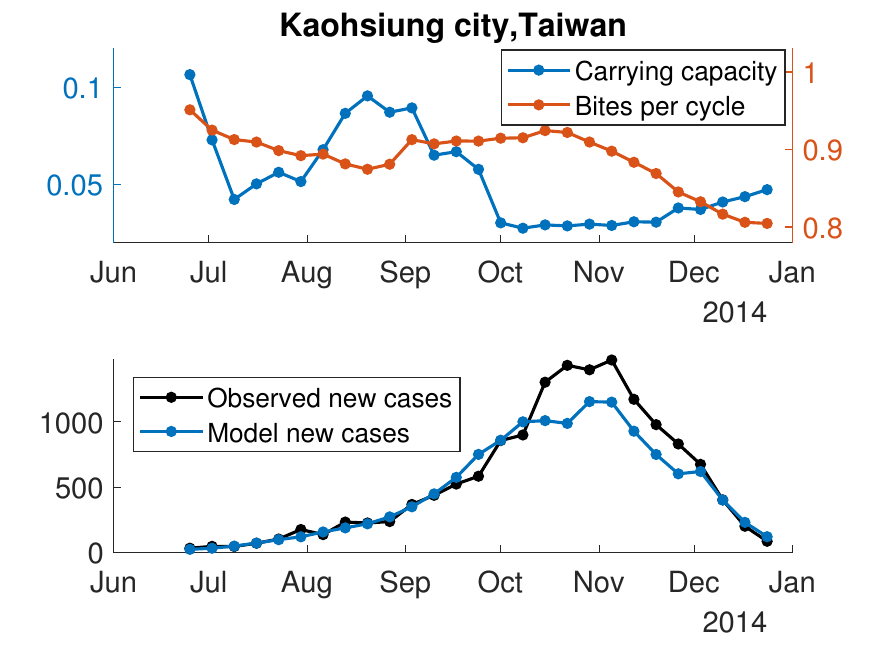} \label{fitlo1}} 
\caption{This figure shows the results of fitting observed human dengue cases to the non-Markovian transmission model described in Section \ref{strm} for two different locations. The unknown model parameters, which are the carrying capacity and the number of bites per gonotrophic cycle, were estimated using the particle filter algorithm for each location (as shown in the plots in the first row). The plots in the second row demonstrate the performance of fitting the model by comparing the observed human cases with the corresponding values obtained from fitting the transmission model.}
\label{fitlo2} 
\end{figure*}

With the exception of the carrying capacity and the number of bites per gonotrophic cycle, all parameters related to dengue transmission model in section \ref{strm} can be found in literature \cite{focks1993dynamic,eisen2014impact,couret2014temperature,yang2011follow,yang2009assessing2,
rossi2014modelling,farnesi2009embryonic,mordecai2017detecting}. The carrying capacity, in general, is a variable that can be impacted by multiple factors, including precipitation and the abundance of water containers within the mosquito habitat. Therefore, constructing a comprehensive model that relates carrying capacity to an assumed set of driving variables requires numerous observations in different locations to cover the variables space. However, data on mosquito abundance are often limited and primarily based on mosquito trap data, which may not accurately reflect the exact number of mosquitoes in a given location. In contrast, dengue transmission data is more readily available, but it is often presented as aggregated numbers of new cases over a wide geographical area. Here, we use dengue new cases time series from 61 different locations \cite{example-website,murphy2020incidence,chaudhry2017dengue,xu2020high,hii2012forecast,
liyanage2016spatial,vasquez2020climate,cuong2011quantifying}, to estimate the carrying capacity and number of bites per gonotrophic cycle. To achieve this, we employ the particle filter algorithm to estimate the unknown transmission model parameters. 

let $X_k=(V_k,H_k,C_k,n_{Bk})$ represent the transmission system at time step $k$, which includes the state vectors for mosquitoes and humans denoted as $V_k$ and $H_k$, as well as the carrying capacity $C_k$ and the number of bites per gonotrophic cycle $n_{Bk}$. To determine the system state at time step $k+1$, we first sample $C_{k+1}$ and $n_{Bk+1}$ from predefined probability distributions. Next, we evolve the mosquitoes and humans vector using the transmission model, taking into account the updated values of the carrying capacity and number of bites per gonotrophic cycle, $C_{k+1}$ and $n_{Bk+1}$. This enables us to obtain a sample for the transmission system at time step $k+1$, denoted as $X_{k+1}=(V_{k+1}, H_{k+1}, C_{k+1}, n_{Bk+1})$. To evaluate the validity of a transmission system sample for time step $k+1$, we compare the number of new human dengue cases derived from $X_{k+1}$ with the observed new cases. In doing so, we assume that the discrepancy between the two values follows a predefined probability density function that peaks at zero. Consequently, we can assign a weight to each sample, which reflects its validity. If we draw a set of such samples, we obtain a vector of weights that we normalize it and thereby assign a probability to the each sample in the set. By selecting the sample set elements according to their assigned probabilities, we obtain an updated set of samples that represents the distribution for the transmission system at time step $k+1$. The particle filtering procedure we described here provides an approximate numerical distribution for the system state, considering a series of observations, i.e., $p(X_{k+1}|n_{k+1},n_{k},\cdots,n_1)$, where $n_{i}$ is the observed new cases at time step $i$. However, in our calculation, we adopt the results of backward particle filter smoothing, which provide a distribution for the trajectory of system state,  formally denoted as $p(X_{k+1},X_{k},\cdots, X_{1} |n_{k+1},n_{k},\cdots,n_1)$. In figure \ref{fitlo2} we have shown the estimated marginal distribution means for the carrying capacity per human, number of bites per gonotrophic cycle and the fitted number of human new dengue cases for two different locations. For these estimations, in the particle filter algorithm, we draw the value of the carrying capacity at step $k+1$, denoted as $C_{k+1}$, from a normal distribution. The mean and standard deviation of this distribution are equal to the carrying capacity at step $k$, denoted as $C_k$, and 50 percent of $C_k$, respectively. We draw the value of $n_{Bk+1}$ in a similar manner except we use 5 percent of $n_{Bk}$ as the standard deviation of the normal distribution. We opt to draw the carrying capacity value from a broad distribution due to its susceptibility to significant impacts from factors such as rainfall and other environmental conditions. 

To investigate the relationship between carrying capacity and precipitation, we fitted the new human case numbers from 61 locations and generated estimated values similar to those shown in Figure \ref{fitlo2}. In the left panel of Figure \ref{fitap}, we plotted the estimated carrying capacity values versus the 2-week moving average precipitation. To provide further insight, the right panel of Figure \ref{fitap} displays the average values of the estimated carrying capacity within precipitation bins of width $0.002$ m, indicated by a cross marker. This plot illustrates that the expected value of carrying capacity increases with low precipitation amounts, but it decreases with large amounts of precipitation. Additionally, we performed fittings of different distributions to the carrying capacity data within each precipitation bin, and we observed that the inverse Gaussian distribution provided a better fit to the data. The inverse Gaussian distribution is a distribution with two parameters that has support on the positive real numbers $(0,\infty)$. If a random variable $C$ follows an inverse Gaussian distribution, it is denoted as $C\sim IG(\mu,\lambda)$, where $\mu$ and $\lambda$ are the distribution parameters. The probability density function for $C$ can be expressed as:
\[
f(c|\mu,\lambda)=\sqrt{\frac{\lambda}{2\pi c^3}} \exp{\left[ -\frac{\lambda(c-\mu)^2}{2\mu^2c}\right]}. 
\]
To fit a statistical model to the carrying capacity data presented in the left panel of Figure \ref{fitap}, we made the assumption that the carrying capacity at each precipitation value follows an inverse Gaussian distribution with the parameters $\mu$ and $\lambda$ that are continuous functions of precipitation. Since the expected value of carrying capacity has different trends for low and high levels of precipitation, we considered the following functions for the model parameters:
\[ \mu(p) = \left\{ \begin{array}{ll}
         a_0+a_1 p+a_2 p^2 & \mbox{if $p < p_0$,}\\
        \alpha_0+e^{-\alpha_1 p}+\alpha_2 & \mbox{if $p \geq p_0$,}\end{array} \right. \] 
 and
 \[ \lambda(p) = \left\{ \begin{array}{ll}
         b_0+b_1 p+b_2 p^2 & \mbox{if $p < p_0$,}\\
        \beta_0+\beta_1 p+\beta_2 p^2 & \mbox{if $p \geq p_0$.}\end{array} \right. \] 
Here, $p$ represents precipitation, and $p_0$ is a threshold value. Although the model appears to have 13 parameters, namely $a_0, a_1, a_2, \alpha_0, \alpha_1, \alpha_2, b_0, b_1, b_2, \beta_0, \beta_1, \beta_2,$ and $p_0$, we can express two of the parameters, such as $\alpha_0$ and $\beta_0$, in terms of the remaining parameters due to the continuity of the functions at $p_0$. 
Using Gibbs sampling we fitted the the statistical model explained above to the carrying capacity data shown in the left panel of figure \ref{fitap}. Since we utilized Bayesian inference to fit the model, we are able to derive marginal posterior distributions for the parameters $\mu$ and $\lambda$ at each precipitation value. In the right panel of Figure \ref{fitap}, the black solid line represents the mean of the marginal posterior distribution of $\mu$, while the dotted red curve represents the $25$th and $75$th percentiles. For a random variable with an inverse Gaussian distribution $IG(\mu,\lambda)$, the expected value is equal to the parameter $\mu$. Therefore, the black solid line in the figure represents the expected value of the carrying capacity. Although we might expect that increasing precipitation leads to an elevated carrying capacity, our results show that this is not always true. However, further data collection is needed to assert this relationship with a high certainty. One explanation for the lower carrying capacity in high precipitation environment might be due to the fact that \textit{Aedes aegypti} is an urban mosquito, and their breeding sites may become flooded during periods of very high precipitation.  
 
\begin{figure*}[t]
\centering
  \subfloat[]{ \includegraphics[width=0.45\columnwidth]{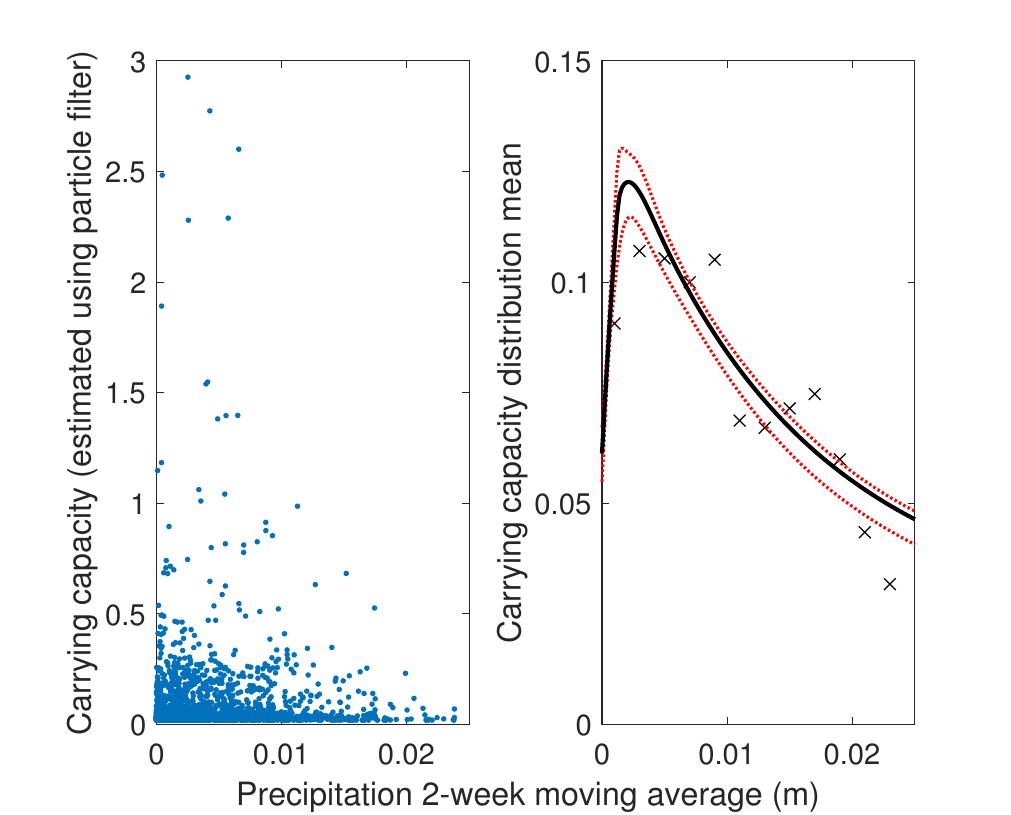} \label{fitap} }%
   \subfloat[]{ \includegraphics[width=0.42\columnwidth]{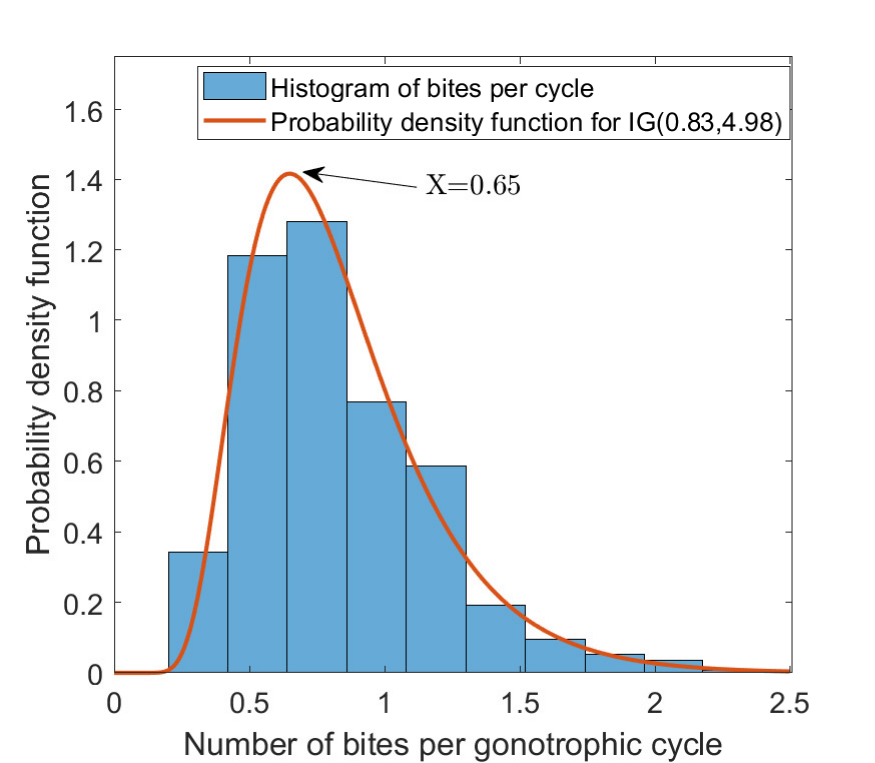}\label{fitbp}}   
   \caption{The left panel of figure (a) shows the estimated carrying capacity values versus the 2-week moving average precipitation. Cross markers in the right panel of figure (a) display the average values of the estimated carrying capacity within precipitation bins of width $0.002$ m, and the solid black curve shows the estimated mean of the inverse Gaussian model fitted to the carrying capacity values in the left panel. The dotted red curves represent the 25th and 75th percentiles of the estimated mean. In plot (b), we have shown the histogram of all the estimated values for bites per gonotrophic cycle, with the density function for the fitted inverse Gaussian distribution shown by the red curve.}  
\label{fit}
\end{figure*}

Fitting the new human dengue cases in 61 locations to the spreading model also provided estimated values for the number of bites per gonotrophic cycle, similar to those shown in figures \ref{fitlo2}. A histogram of all the estimated values for bites per gonotrophic cycle is presented in figure \ref{fitbp}. Additionally, we found that the inverse Gaussian distribution with the mean of 0.83 provides the best fit to the estimated values. Although this value is close to one blood meal per cycle, the lower expected value suggests that not all blood meals are successful in the transmission of the dengue virus. 
\subsection{Dengue risk map} \label{riskmapsec}
\begin{figure*}[t]
\centering
  \subfloat[]{ \includegraphics[width=0.45\columnwidth]{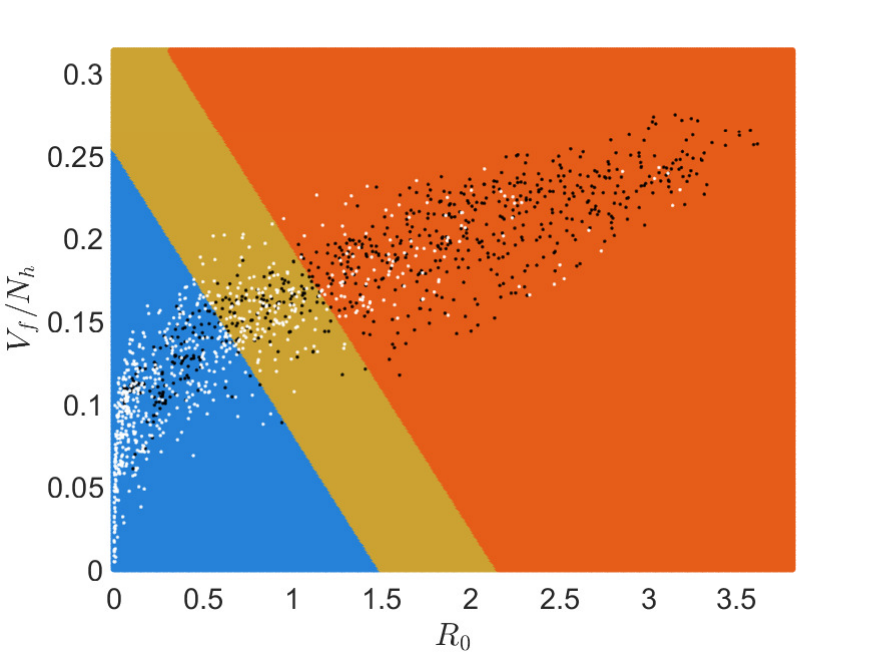} \label{riskmodela} }%
   \subfloat[]{ \includegraphics[width=0.45\columnwidth]{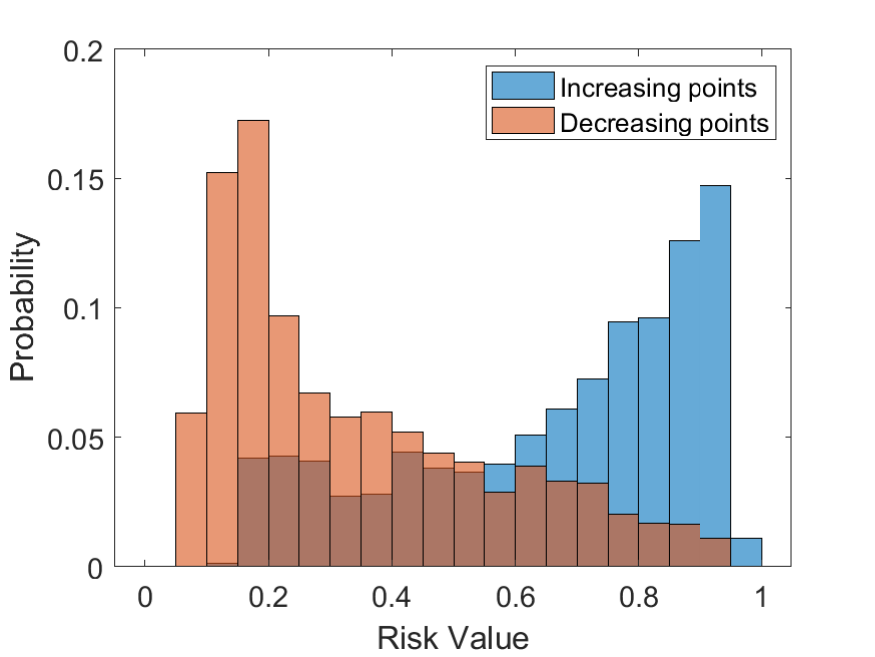}\label{riskmodelb}}   
   \caption{Plot(a) displays random samples of the weeks with increasing and decreasing dengue cases in different locations as black and white dots, respectively. The $R_0$ and $V_f$ values in the plot correspond to the 30-day moving average of the estimated reproduction number and mosquito population for the respective week. Furthermore, in this plot, the region shaded in red has an outbreak probability greater than $0.6$ according to the trained model described in Section \ref{riskmapsec}. Meanwhile, the outbreak probability in the blue-shaded area is less than $0.4$. Plot(b) displays the distribution of calculated risk values for the training data, which includes weeks with both increasing and decreasing human dengue cases.}  
\label{riskmodel}
\end{figure*}

By definition, the dengue reproduction number, given by the equation \ref{repro}, provides a threshold that determines whether the virus can spread in a healthy population and result in an increasing number of new cases or not. While some studies \cite{mordecai2017detecting} have utilized the value of $R_0$ as a risk factor, the application of this parameter necessitates certain considerations. First, the reproduction number is derived under the assumption of a constant temperature, providing an epidemic threshold based on this specific condition. Second, the calculation of $R_0$ requires an estimation of the number of female mosquitoes, along with its variations influenced by precipitation and temperature. To address these issues, we only use reproduction number as a predictive variable, coupled with an estimation of the mosquito population within a numerical model for assessing the dengue risk factor. Furthermore, we incorporate the model developed for the fluctuation in carrying capacity, as outlined in section \ref{carmod}, to estimate the variation in mosquito population. This variation plays a pivotal role in the reproduction number given by equation \ref{repro}. 
To create a numerical model for the dengue transmission risk factor, we initially manually labeled the dengue epidemic curves from various locations, \cite{example-website,murphy2020incidence,chaudhry2017dengue,xu2020high,hii2012forecast,
liyanage2016spatial,vasquez2020climate,cuong2011quantifying}, with either a one or zero, indicating the weeks with increasing or decreasing cases. Subsequently, we calculated the corresponding values of the reproduction number and mosquito population for the labeled weeks. This calculation relied on temperature and precipitation data for the respective locations, which we obtained from the ERA5 global reanalysis data \cite{hersbach2020era5}. ERA5 provides daily data with a resolution of 0.25 degrees in longitude and latitude and is accessible through the Copernicus Climate Change Service \cite{era5}. In figure \ref{riskmodela}, we have shown random samples of the weeks labeled with zero and one as white and black dots, respectively. Furthermore the x and y values of a labeled week in the plot correspond to the 30-day moving average of the estimated reproduction number and mosquitoes population for the corresponding week. To train a numerical model for dengue transmission risk, we fed the labeled data into a feedforward neural network, which assigns a value between zero and one to points on the two-dimensional surface of reproduction number and mosquito population. This value can be interpreted as the probability of dengue outbreak or having increasing number of cases. It is possible to train neural network models with numerous nodes in the layers, effectively dividing the two-dimensional surface into high and low-risk areas in a nonlinear manner. However, to avoid overfitting, we used a neural network with only one node, which effectively corresponds to a logistic regression model. In figure \ref{riskmodela}, we have depicted the classification resulting from the trained model. The region with assigned values greater than $0.6$ is colored in red, while the region with values smaller than $0.4$ is colored in blue. The slope of the boundaries of the regions in this figure indicates that the $R_0$ value is not the sole factor determining the evaluation of transmission risk. This might be due to the fact that the $R_0$ expression is derived for constant temperature and includes some terms that can experience significant fluctuations with temperature. To further demonstrate the fitting performance, in figure \ref{riskmodelb}, we have shown the distribution of the calculated risk values for the training data points. Although the distributions for the increasing and decreasing points are not completely separated, they are concentrated in the regions of high and low-risk values, respectively. 

\begin{figure*}[t]
\centering
\subfloat[]{ \includegraphics[width=0.5\columnwidth]{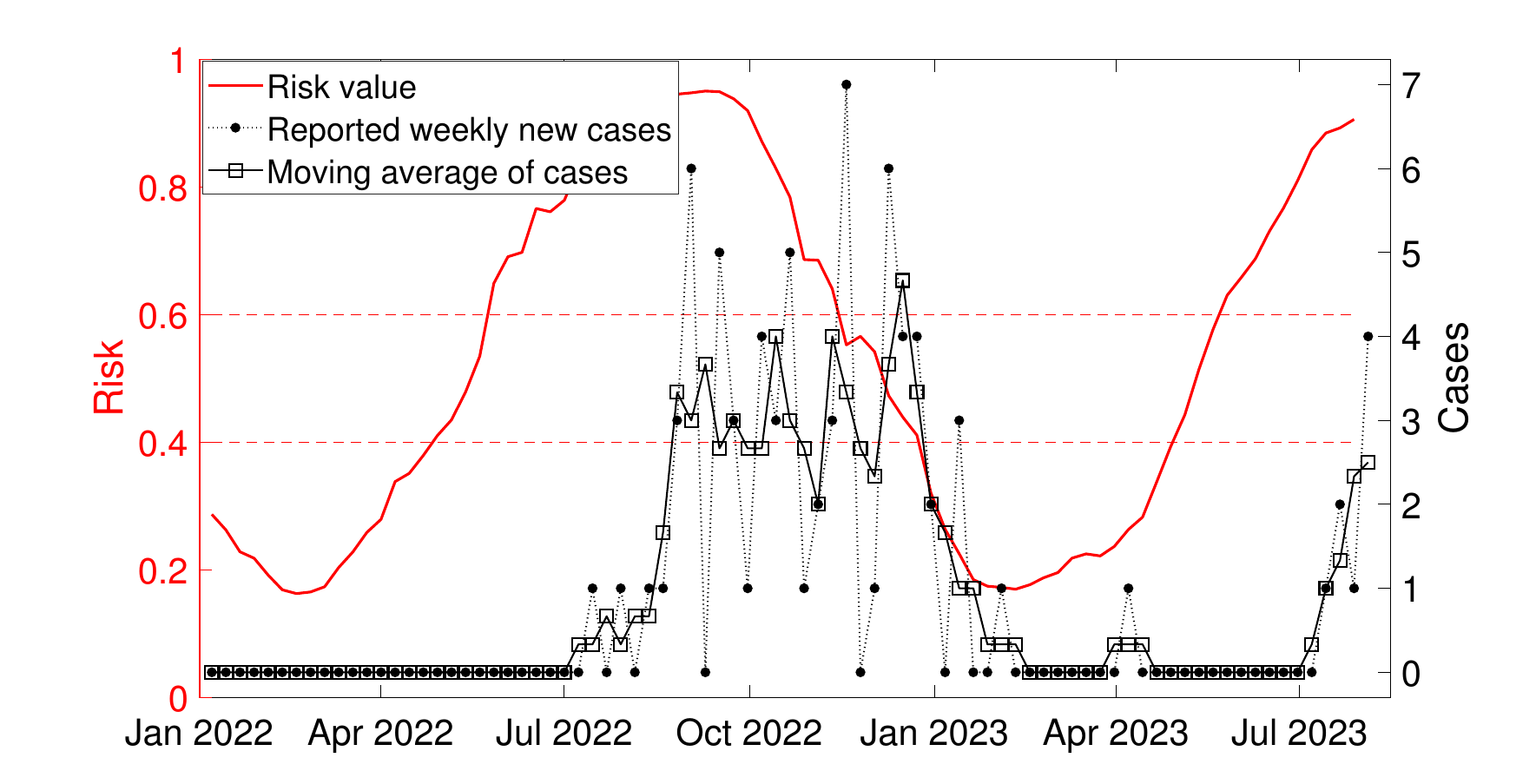}} 
\subfloat[]{ \includegraphics[width=0.5\columnwidth]{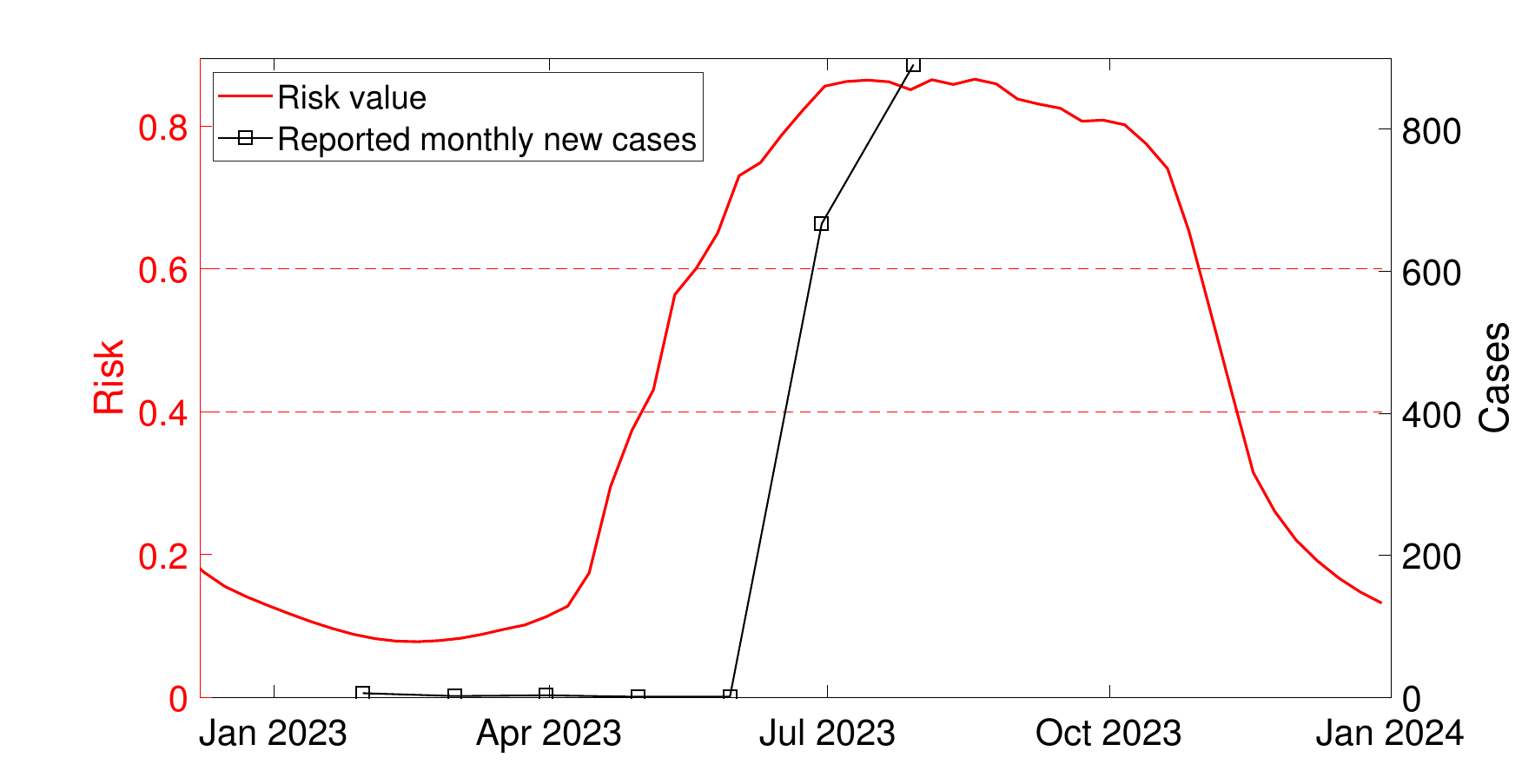}}
\caption{The plots show human new dengue cases along with the calculated transmission risk for Florida, USA (a), and Sunsari, Nepal (b).}
\label{floridav}
\end{figure*}

To demonstrate the utility of the risk model, in Figure \ref{floridav}, we have plotted the human new dengue cases along with the calculated transmission risk for Florida, USA (a), and Sunsari, Nepal (b). We can observe that local transmission occurs during high-risk periods. For instance, in Florida, between January and March 2023, when the risk is low, the number of new cases exhibits a declining trend.

\begin{figure*}[t]
\centering
  \subfloat[]{ \includegraphics[width=0.45\columnwidth]{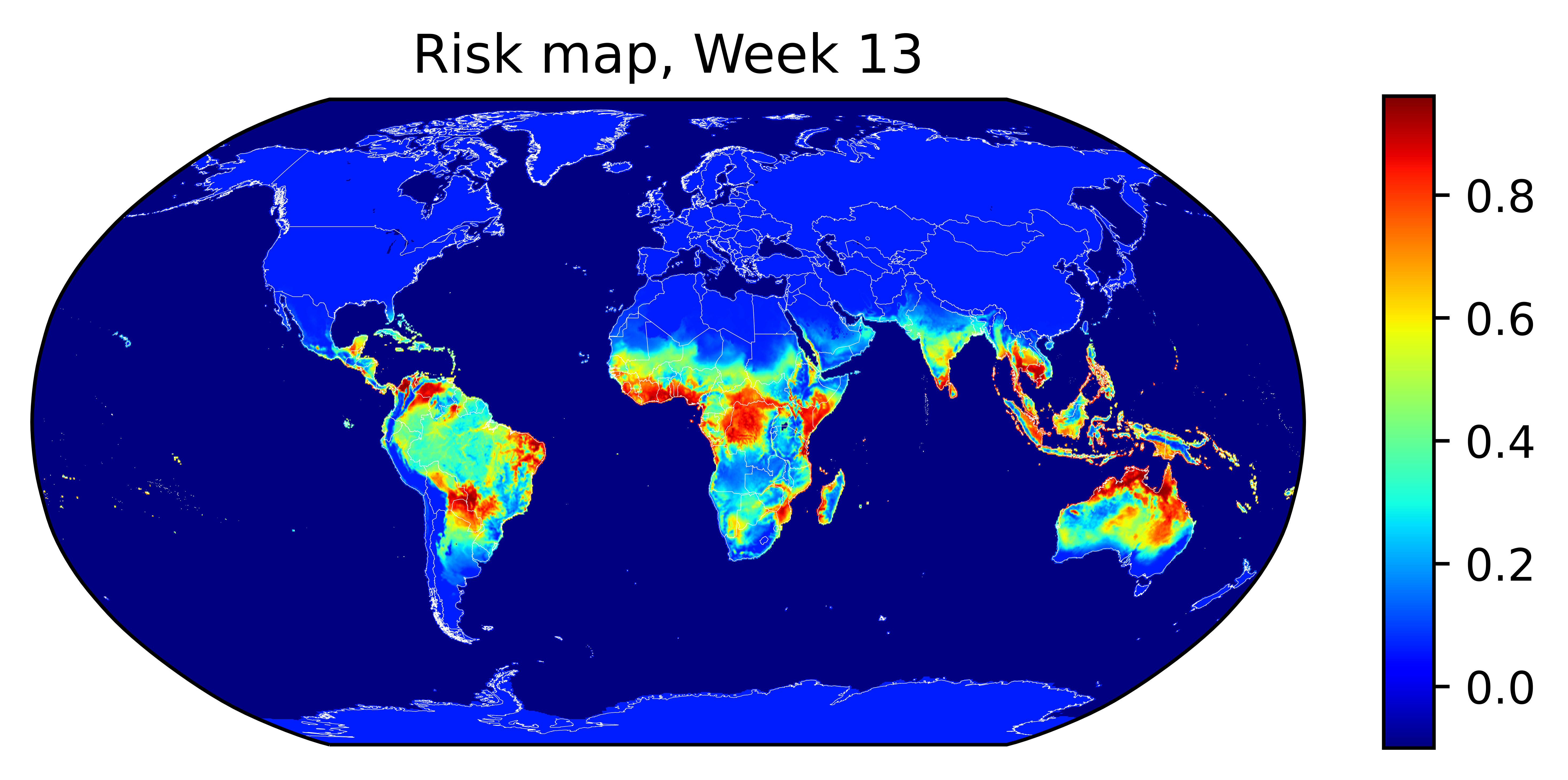} \label{fitap} }  
     \subfloat[]{ \includegraphics[width=0.45\columnwidth]{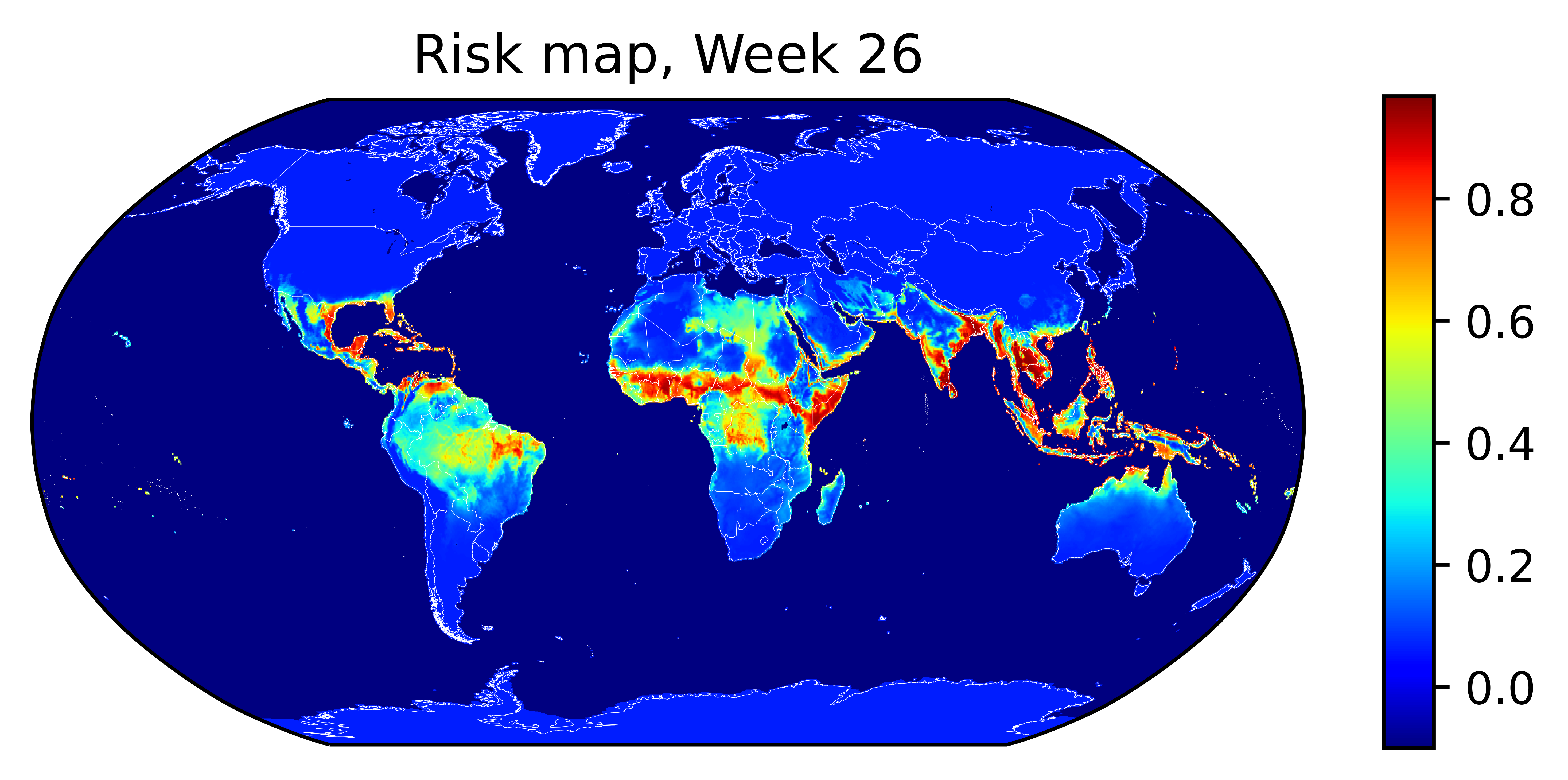} \label{fitap} }\\
      \subfloat[]{ \includegraphics[width=0.45\columnwidth]{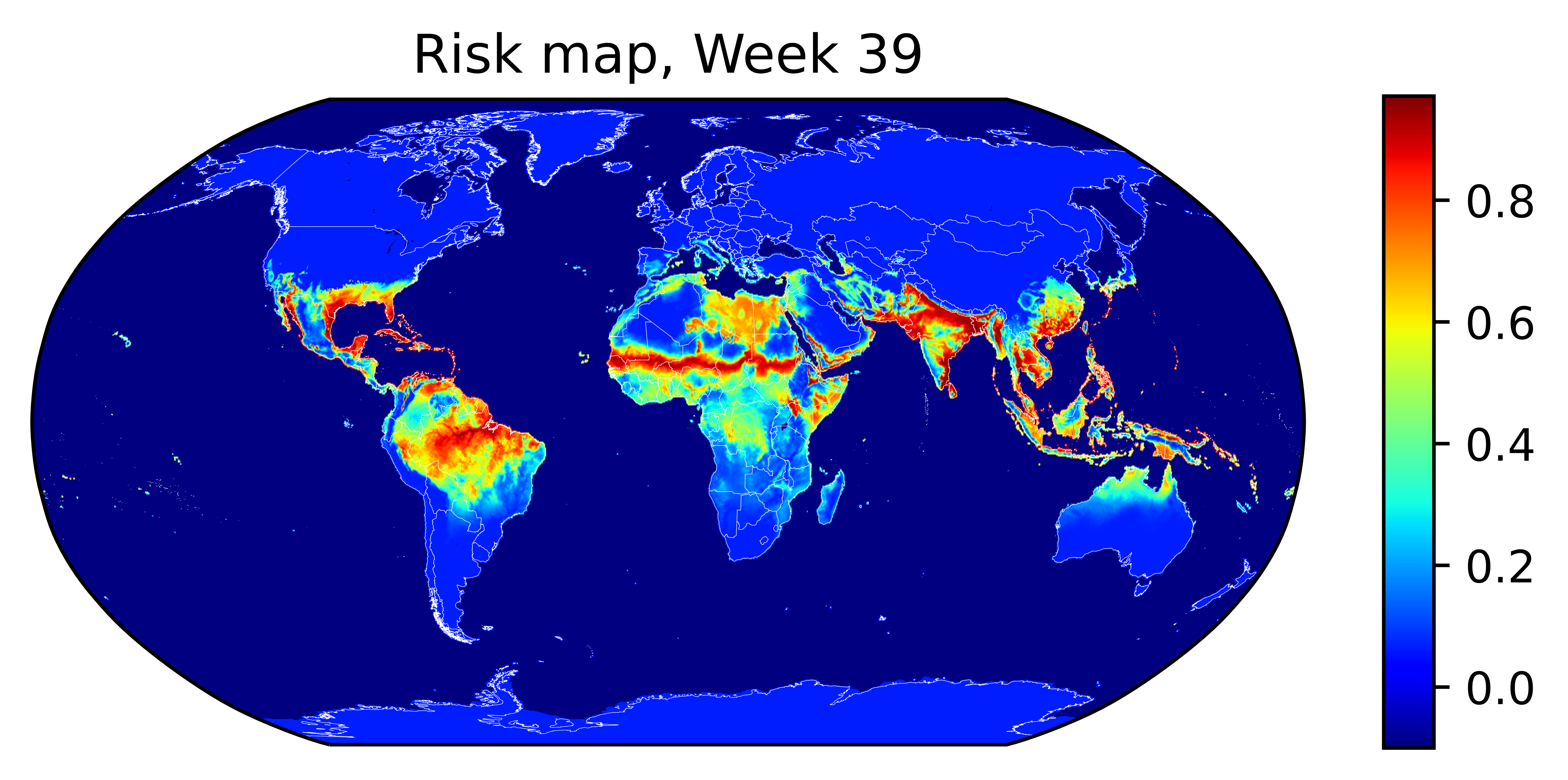} \label{fitap} }
      \subfloat[]{ \includegraphics[width=0.45\columnwidth]{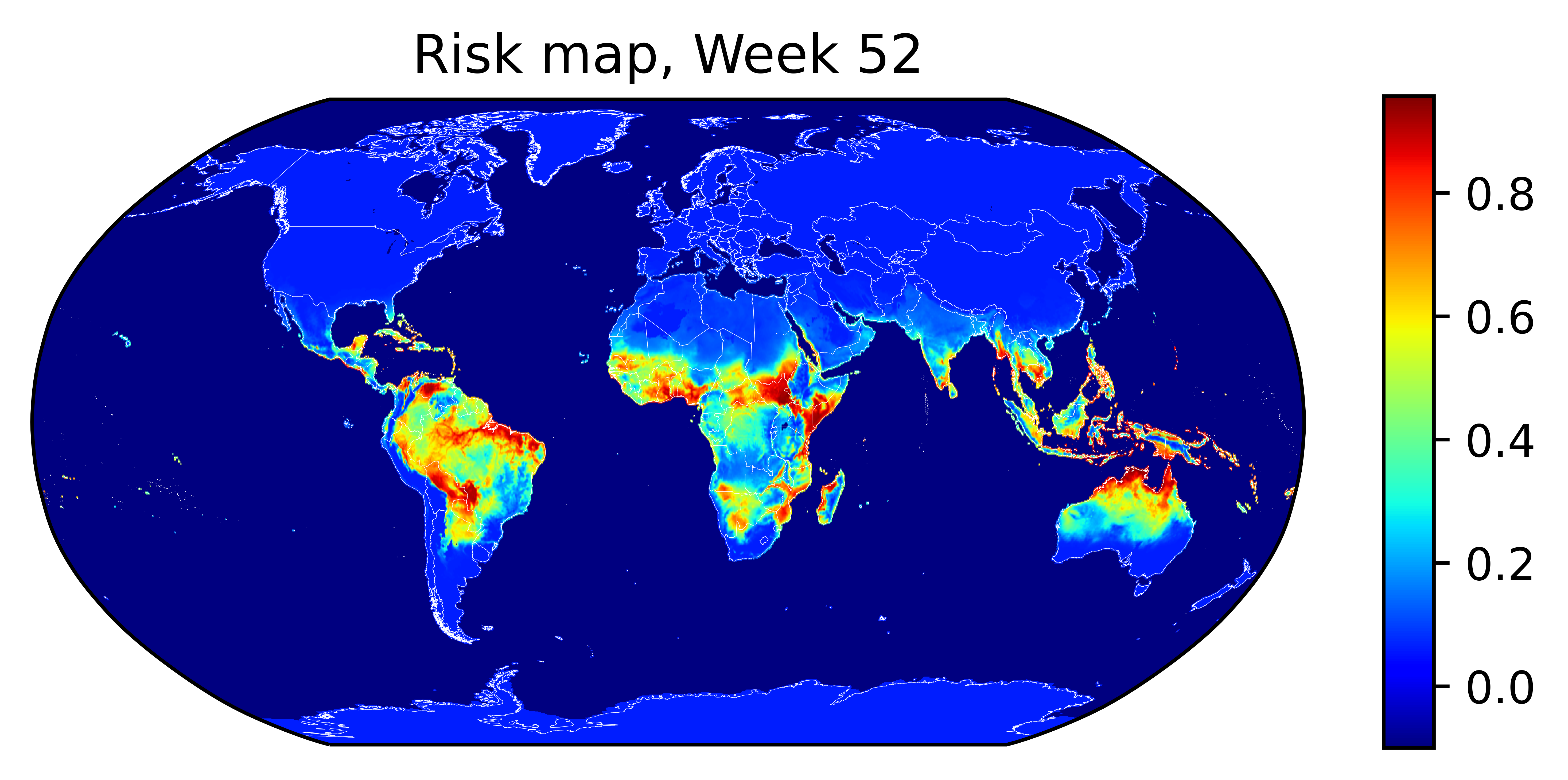} \label{fitap} }  
   \caption{The figures show global dengue transmission risk for four different weeks in the year 2022.}  
\label{worldrisk}
\end{figure*}

Transmission risk values can be determined for any location, provided access to local temperature and precipitation data. To accomplish this, we obtained temperature and precipitation data for the year 2022 from the Copernicus Climate Change Service \cite{era5}, which offers daily global data with a resolution of $0.25$ degrees in both latitude and longitude. By utilizing the daily average data, we constructed the daily risk map. Figure \ref{worldrisk} presents risk maps at four distinct time points throughout the year. When interpreting these maps, it's important to consider that our calculations assume the presence of \textit{Aedes aegypti} mosquito and the dengue virus in a given location, and subsequently assess the likelihood of mosquito survival and dengue transmission. As a result, these maps can indicate the suitability of various locations for the survival of the mosquito population.

\section{Conclusions}
In this study, we developed a new model for the \textit{Aedes aegypti} mosquito life cycle, the primary vector for Dengue. We employed this model to create a global dengue risk map with daily temporal resolution and a spatial resolution of 0.25 degrees in longitude and latitude, which is available online \cite{picturee,yi2023picturee}.
In our model, we assume that the daily development of mosquitoes at each life-cycle phase depends on the average daily temperature. We accumulate daily development values until the mosquitoes' development is complete, at which point they transition to the subsequent life history stage or phase. Our calculations indicate that in cases of significant temperature fluctuations throughout a year, particularly with prolonged periods of lower temperatures, the traditional non-homogeneous ODEs with temperature-dependent rates predicts an earlier increase in the mosquito population compared to the output of our model. This discrepancy arises from the inherent assumption within traditional non-homogeneous ODEs, as demonstrated in section \ref{odesec}, that the transition to a new lifecycle stage occurs when the development reaches a value that is distributed exponentially with a mean of one. This leads to earlier mosquito emergence compared to our model, where the transition occurs upon completion of development. This difference becomes more pronounced in environments with lower temperatures, which result in longer development times. 

Although the mosquito life cycle is non-Markovian process and is therefore best described by integral equations, we have shown that it is possible to reduce the master equations to ODEs by assuming the general class of phase-type distribution for the random variables associated with the development values at the transition times. This is particularly useful because it enabled us to analyze the system using well-established methods developed for ODEs and derive the corresponding reproduction number of dengue and the mosquito's offspring number. Furthermore, the reduction to ODEs enabled us to utilize the particle filter algorithm for estimating the environmental carrying capacity and biting rate by fitting the outbreak data to the dengue transmission model. This is significant because the particle filter approach is developed for Markovian systems. 

The developed risk map relies on historical outbreak data for the carrying capacity model and outbreak risk assessment itself. While we utilize outbreak data from various global locations, there is room for enhancing both the carrying capacity and risk models. Our model, which links carrying capacity to precipitation, demonstrates a decrease in carrying capacity when the 2-week moving average of daily precipitation exceeds 1 cm. This reduction could be attributed to excess precipitation washing out larvae from habitats when they overflow or if pools and ponds are scoured clean by excessive flow. However, further data collection is needed to solidify the relationship between carrying capacity and precipitation. Upon comparing the resultant risk values with actual human dengue case data from Florida and Nepal, we observed a strong correlation between the risk model and the case data. This suggests that the risk assessment tool has practical utility in guiding intervention measures, including mosquito control strategies. Additionally, the risk map assigns a high risk to locations like southern Iran, which have only recently been invaded by \textit{Aedes} mosquitoes. This observation holds significance because our risk model relies solely on precipitation and temperature data. This indicates that the risk model has the potential to predict areas where \textit{Aedes aegypti} mosquito populations could establish themselves, even in locations where they are not currently present.

\ifCLASSOPTIONcompsoc
  \section*{Acknowledgments}
\else
  \section*{Acknowledgment}
\fi
The project was sponsored by the Department of the Army, U.S. Army Contracting Command, Aberdeen Proving Ground, Natick Contracting Division, Ft Detrick, MD.
Any opinions, findings, and conclusions or recommendations expressed in this material are those of the authors and do not necessarily reflect the position or the policy of the Government, and no official endorsement should be inferred.

\ifCLASSOPTIONcaptionsoff
  \newpage
\fi

\bibliographystyle{IEEEtran}
\bibliography{Refrences}

%




\end{document}